\pdfoutput=1
\documentclass[a4paper,11pt]{article}
\usepackage{amsmath,amssymb,amsfonts}
\usepackage[multiple,stable]{footmisc}
\usepackage[amsmath,amsthm,thmmarks]{ntheorem}
\allowdisplaybreaks[4]
\usepackage[noconfig]{refstyle}
\usepackage[dvipsnames]{xcolor}
\usepackage[hyperfootnotes=false, linktocpage=true, colorlinks, citecolor=blue, linkcolor=blue, urlcolor=Maroon]{hyperref}
\usepackage{array,tabularx,booktabs,geometry,subfigure,graphicx,longtable}
\usepackage[bf]{caption}
\usepackage{tikz}
\usepackage[all]{xy}
\usetikzlibrary{shapes,arrows}
\tikzstyle{block} = [rectangle, draw, text width=7em, text centered, rounded corners, minimum height=3em]
\usepackage{graphicx,color}
\usepackage{cite}
\usepackage{tikz}
\usepackage{tikz-cd}
\usetikzlibrary{cd}

\let\eqref=\relax
\newref{eq}{name={eq.~},Name={Eq.~},names={eqs.~},Names={Eqs.~},rngtxt={-},refcmd=(\ref{#1})}
\newref{f}{name={footnote~},Name={Footnote~},names={footnotes~},Names={Footnotes~}}
\newref{app}{name={appendix~},Name={Appendix~},names={appendixes~},Names={Appendixes~}}
\newref{tab}{name={table~},Name={Table~},names={tables~},Names={Tables~}}
\newref{ch}{name={chapter~},Name={Chapter~},names={chapters~},Names={Chapters~}}
\newref{sec}{name={section~},Name={Section~},names={sections~},Names={Sections~}}
\newref{fig}{name={figure~},Name={Figure~},names={figures~},Names={Figures~}}
\numberwithin{equation}{section}
\newcommand{\eref}[1]{(\ref{#1})}

\newcommand{\ba}{\begin{array}}
\newcommand{\ea}{\end{array}}

\newcommand{\IP}{\mathbb P}

\newcommand{\be}{\begin{equation}}
\newcommand{\ee}{\end{equation}}
\newcommand{\bea}{\begin{equation}\begin{aligned}}	
\newcommand{\eea}{\end{aligned}\end{equation}}		

\newcommand{\iddots}{\mathinner{\mkern2mu\raise1pt\hbox{.}\mkern2mu \raise4pt\hbox{.}\mkern2mu\raise7pt\hbox{.}\mkern1mu}}

\providecommand{\id}{\leavevmode\hbox{\small$\mathrm{1}$\kern-3.8pt\normalsize$\mathrm{1}$}}
\def\fnote#1#2{\begingroup\def\thefootnote{#1}\footnote{#2}
     \addtocounter{footnote}{-1}\endgroup}

\begin{document}

\vspace{1cm}

\title{
       \vskip 40pt
       {\huge \bf Heterotic Complex Structure Moduli Stabilization for Elliptically Fibered Calabi-Yau Manifolds}}

\vspace{2cm}

\author{Wei Cui${}^1$, Mohsen Karkheiran${}^{2}$}
\date{}
\maketitle
\begin{center} {\small ${}^1${\it Physics Department, Robeson Hall, Virginia Tech, Blacksburg, VA 24061, USA}\\
\small${}^2$Center for Theoretical Physics of the Universe,
Institute for Basic Science, Daejeon 34051, South Korea}\\
\fnote{}{cwei@vt.edu, mohsenkar@ibs.re.kr}
\end{center}

\begin{abstract}
\noindent
Complex structure moduli of a Calabi-Yau threefold in $N=1$ supersymmetric heterotic compactifications can be stabilized by holomorphic vector bundles. 
The stabilized moduli are determined by a computation of Atiyah class. 
In this paper, we study how this mechanism work in the context of elliptically fibered Calabi-Yau manifolds where complex structure moduli space contains two kinds of moduli, ones from base and ones from fibration.  
With spectral cover bundles, we find three types of situations when holomorphicity of bundles is determined by algebraic cycles supported on special choice of complex structure, which allows us to stabilize both of these two moduli.
We present concrete examples for each type and develop practical tools to analyze the stabilized moduli. 
Finally, by checking the holomorphicity of the four-flux and/or local Higgs bundle data in F-theory, we briefly study the dual complex structure moduli stabilization scenarios.

\end{abstract}

\thispagestyle{empty}
\setcounter{page}{0}
\newpage

\tableofcontents

\section{Introduction} \label{sec1}

Compactification of $E_8 \times E_8$ heterotic string \cite{Candelas:1985en,Green:1987mn} 
on a Calabi-Yau threefold with a vector bundle is a promising approach to derive the Standard Model of the particle physics in four-dimensional spacetime \cite{Bouchard:2005ag,Braun:2005zv,Braun:2005bw,Braun:2005ux,Anderson:2009mh,Braun:2011ni,Anderson:2011ns,Anderson:2012yf,Anderson:2013xka}.
To preserve supersymmetry, the bundle should be holomorphic, slope zero and poly-stable. The fluctuations of the Calabi-Yau's complex structure and Kahler structure and the bundle's fluctuations appear as moduli particles in the effective theory.
The holomorphic property of the bundle has been used to stabilize complex structure moduli in these models \cite{Anderson:2010mh,Anderson:2011ty,Anderson:2011cza,Anderson:2013qca,Gray:2019tzn}. 
The idea is that only some of the deformations of complex structure are consistent with the supersymmetry. 
The others will force the bundle to become non-holomorphic. 
This will generate an non-trivial F-terms for such deformations in the effective theory and the complex structure moduli corresponding to them are stabilized. As shown in \cite{Anderson:2011cza}, this mechanism together with a similar one due to bundle poly-stability \cite{Sharpe:1998zu,Anderson:2009sw,Anderson:2009nt,Anderson:2010ty,Anderson:2010tc} is an effective way to stabilize geometric moduli in heterotic Calabi-Yau compactification.

Deformations of complex structure preserving bundle holomorphy are described by Atiyah class \cite{atiyah}.
The computation of it is in general difficult. However, for holomorphic bundles in which the necessary ingredients in the construction have a complex structure dependence, these deformations can be calculated efficiently by a ``bottom up'' approach. 
For example, extension bundles \cite{Braun:2005zv} are defined by specifying an element in Ext group. 
As shown in \cite{Anderson:2010mh,Anderson:2011ty}, Ext group can jump in dimension for special choice of complex structure, which stabilize some of complex structure moduli locally on special locus in moduli space.
Instead of Atiyah class, the stabilized moduli can be identified quickly by a computation of ``jumping'' Ext. 
As a result, the ''bottom up'' approaches are equivalent to calculation of Atiyah class and are efficient tools for complex structure moduli stabilization. 



In this work, we will focus on Calabi-Yau threefold with an elliptic fibration\footnote{The Calabi-Yau manifolds considered here are all in the Weierstrass form. This paper's ideas will be valid for any elliptic fibration, as long as there is at least one holomorphic section. However, when all of the sections are rational, one will need more careful analysis \cite{Anderson:2019agu,Anderson:2019axt}.} \cite{Gross:1993fd,Constantin:2016xlj,Anderson:2016ler,Anderson:2017aux,Anderson:2018kwv,Huang:2018esr,Huang:2018vup,Huang:2019pne} and study how its complex structure moduli can be stabilized by bundle holomorphy. 
Bundles on elliptically fibered Calabi-Yau threefold can be constructed by spectral cover \cite{Friedman:1997yq,Friedman:1997ih,Anderson:2019agu,Anderson:2019axt}, which can be mapped to the extension or monad bundles by a Fourier-Mukai transformation. The reason to work with spectral cover bundle is that
poly-stability, a property that is difficult to check for other constructions, is straightforward to check in terms of spectral data at least as long as the spectral cover is smooth. Furthermore, for technical reasons, the models considered so far are restricted to $SU(2)$ bundles, whereas working with spectral data does not restrict the rank of the bundle.\footnote{It is interesting to see how many of the moduli fields can be stabilized in the phenomenologically interesting models using the methods presented in this paper. However, we only explain the general ideas here and leave such questions for a future project.} Besides, since spectral data specify the Heterotic models, the Heterotic model's moduli stabilization process can be translated to an associated process in the dual F-theory model. Hence the relation with moduli stabilization scenarios in F-theory/type II models will be manifest.

Complex structure moduli of an elliptically fibered Calabi-Yau threefold $X$ can be either the complex structure moduli of the base $B$ or complex structure moduli of the fibration $F$. 
As we will see later, both of them can be stabilized using holomorphic bundles. Specifically, we find three types of approaches to do that but, from the perspective of spectral cover bundles, they all correspond to emergent algebraic cycles in $X$. 
In more technical terms, we look for the Noether-Lefschetz loci in the complex structure moduli and specify the spectral data so that the associated vector bundle stays holomorphic on these loci.
For type I case, we consider an $SU(2)$ extension bundle with Ext group jumping on special locus in complex structure moduli space of $B$. By Fourier-Mukai transformation, elements in jumping Ext group corresponds to new algebraic curves in the spectral cover. One can observe that, variations of complex structure away from the special locus in moduli space will make algebraic curve disappear and spectral cover is not algebraic anymore. Hence the corresponding vector bundle becomes non-holomorphic and the associated complex structure moduli of $B$ are stabilized.

In type II approach, we find that Picard group of $B$ jumps when the complex structure of the base takes some ``special'' form. 
Suppose the holomorphic bundle is defined with the new algebraic cycles in ``jumping'' Picard group. 
Then, variations of complex structure that does not preserve this ``special'' form of complex structure will make the bundle non-holomorphic and the corresponding moduli are stabilized. 
Besides complex structure moduli of the base, holomorphic bundles can also stabilize moduli of the fibration. 
The idea of type III approach is similar with type II one, in which taking a ``special'' complex structure gives new algebraic cycles in the base. 
The difference is that this time, the algebraic curves appear in the total space of the fibration. Suppose the spectral data are defined in term of these emergent algebraic cycles.
Then, variations of complex structure away from the ``special'' form will make algebraic cycles non-algebraic and hence the bundle becomes non-holomorphic and the associated moduli are stabilized. 
Notice that the last two approaches does not require the bundle is constructed as an extension and can be applicable to more general bundles. 
We give explicit examples for all three approaches, and similar to ``jumping'' Ext, we develop practical tools to identify variations of complex structure that does not preserve the ``special'' form used above and thus obtain the stabilized moduli in type II and type III approaches.

The image of these moduli stabilization scenarios are also interesting. We will show the type I example (which is the usual stabilization example by rank $2$ bundles) the F-theory complex structure moduli is stabilized by a four-flux which in some way is related to the isolated vertical fibers of the spectral cover. In type II example the flux that is derived from the spectral line bundle is responsible for the moduli stabilization. Finally the four-flux in type II example is actually the horizontal algebraic four-flux constructed in \cite{Braun:2011zm}. A recent work on complex structure moduli stabilization also used the idea of algebraic cycles is discussed in \cite{Braun:2020jrx}.

The organization of this paper is as follows: 
In Section \ref{sec2}, we review how the holomorphic property of a vector bundle can be used for complex structure moduli stabilization in Calabi-Yau compactification. 
We study this mechanism in terms of the the Atiyah class and illustrate it with a $SU(2)$ extension bundle.
In Section \ref{sec3}, we introduce three approaches to stabilize the complex structure moduli of an elliptically fibered Calabi-Yau threefold with holomorphic bundles in spectral cover construction. 
In Section \ref{sec4}, we discuss how these approaches in heterotic compactification manifest themselves in F-theory.
In Section \ref{sec5}, we conclude our work.
Some of the technical discussions are presented in the appendices.

\section{The Atiyah Class} \label{sec2}

Let $X$ be a Calabi-Yau threefold with a vector bundle $V$. To preserve $N=1$ supersymmetry, the bundle should have a connection satisfying the following (zero slope) Hermitian Yang-Mills equations,
\begin{equation} \label{hym}
  F_{ab}=F_{{\bar a}{\bar b}}=0,   \; g^{a \bar{b}} F_{a \bar{b}} = 0, 
\end{equation}
where $F$ is the gauge field strength associated with that connection and $a$ and
$\bar{b}$ are holomorphic and anti-holomorphic indices on the Calabi-Yau manifold.
The first equation in (\ref{hym}) implies that the bundle is holomorphic \cite{hartshorne}. It is the property that will be used for moduli stabilization.  
The second equation in (\ref{hym}) means that the bundle is poly-stable \cite{Donaldson:1985zz,duy}. We will always keep $V$ poly-stable when studying variations of complex structure of $X$. 
As shown in \cite{Anderson:2010mh,Anderson:2011ty}, one can do that as long as the bundle is holomorphic.

We start with a supersymmetric vacuum where the initial K\"ahler, complex structure and bundle moduli are chosen to obey the Hermitian Yang-Mills equation in (\ref{hym}). Notice that the initial objects will be denoted with a superscript ``$(0)$'' and their deformation will be denoted with a $\delta$ in front. 
From the deformation theory of compact complex manifolds \cite{atiyah, kodaira, kobayashi,kuranishi,donaldson_def},
the space of simultaneous holomorphic deformations of $X$ and $V$ is measured by $H^{1}({\cal{Q}})$ with $Q$ satisfying the Atiyah sequence 
\begin{equation} \label{atiyahseq}
0 \to V\otimes V^{\vee} \to {\cal Q} \stackrel{\pi}{\to} TX \to 0 \;.
\end{equation}
The extension class is 
\begin{equation} \label{atiyahcalss}
\alpha=[F^{(0)1,1}] \in H^1(V\otimes V^{\vee} \otimes TX^{\vee})
\end{equation}
the $(1,1)$ component of the initial field strength $F^{(0)}$ of the background, referred to as the ``Atiyah class'' \cite{atiyah}. From the long exact sequence of (\ref{atiyahseq}), we have
\begin{equation} \label{H1Q}
0 \to H^1(V\otimes V^{\vee}) \to H^1({\cal Q}) \stackrel{d\pi}{\to} H^1(TX) \stackrel{\alpha}{\to} H^2(V\otimes V^{\vee}) \to \ldots  \;.
\end{equation}
The simultaneous holomorphic deformations space $H^1({\cal Q})$ is given by 
\begin{equation} \label{H1Q2}
H^1({\cal Q})=H^1(V\otimes V^{\vee}) \oplus {\rm Ker} (\alpha) \ . 
\end{equation}
Notice that we have used the fact that $H^0(TX)=0$ for the stable bundle $TX$ in (\ref{H1Q}).

As we can see in (\ref{H1Q2}), all bundle moduli in $H^1(V\otimes V^{\vee})$ belong to $H^1({\cal Q})$ while for complex structure moduli, only those that are in the kernel of the Atiyah class are elements of $H^1({\cal Q})$ and should be regarded as the ``true'' moduli in the effective theory.  
The other ones have non-zero images in $H^2(V\otimes V^{\vee})$, which lead to a non-vanishing field strengths $F^{0,2}$ in the background and make $V$ non-holomorphic. 
Thus, variations of complex structure along those directions are stabilized by a nontrivial F-terms in superpotential.
The number of stabilized moduli is counted by the dimension of ${\rm Im}(\alpha)$. 
From the sequence in (\ref{H1Q}), it is bounded by $h^2(V\otimes V^{\vee})$, so there are at most $h^2(V\otimes V^{\vee})$ complex structure moduli that can be stabilized by the holomorphic bundle.

The above mechanism can be made more precise if we write the elements of $H^{1}(TX)$ explicitly as $\nu=\delta {\mathfrak{z}}^{I} v_{I {\bar{a}}}^{c}$ where $v_I$ are tangent bundle
valued harmonic one-forms and $\delta\mathfrak{z}^I$ are variations of
the complex structure moduli $\mathfrak{z}^I$. 
By the sequence (\ref{H1Q}), elements in the kernel of Atiyah class should satisfy 
\begin{equation} \label{atiyah}
\delta{\mathfrak z}^{I}v^{c}_{I[{\bar a}}F^{(0)}_{|c|{\bar b}]}=D^{(0)}_{[{\bar a}}\Lambda_{{\bar b}]} \,
\end{equation}
where $\Lambda$ is a bundle-valued one-form and $D^{(0)}$ is the covariant derivative with respect to the initial connection $A^{(0)}$. 
Notice that the right hand side of the equation above is an exact 2-form in $H^2(X,V \otimes V^{\vee})$. In fact, the bundle-valued one-form $\Lambda$ can be understood as deformation of the connection $\Lambda_{\bar{b}} = \delta A_{\bar{b}}$. 
Thus, for deformations of complex structure $\delta{\mathfrak z}^{I}$ that are in ${\rm Ker}(\alpha)$, 
there exist a deformation of the connection $\delta A$ such that the the equation (\ref{atiyah}) is satisfied and $V$ can remain holomorphic. However, other deformations will make $V$ non-holomorphic and the associated moduli are stabilized. 

In conclusion, with Atiyah class defined in (\ref{atiyahcalss}), we can determine which deformations of the complex structure in $H^{1}(TX)$ can keep $V$ holomorphic and which are not. 
The number of these independent deformations are given by the dimension of ${\rm Ker}(\alpha)$ and ${\rm Im}(\alpha)$ respectively. 
For most general bundles, they can be computed directly from the long exact sequence in (\ref{H1Q}) known as the ``top down'' approach. 
However, for some classes of holomorphic bundles constructed with ingredients that are only well-defined for special complex structure, there is a quick way to determine the ${\rm Ker}(\alpha)$ and ${\rm Im}(\alpha)$. 
This ``bottom up'' approach is equivalent 
\footnote{
For the rank two vector bundles constructed as extension of a line bundle and its dual, it was proved in \cite{Anderson:2011ty} that the Atiyah class computation is rigorously equivalent to an analysis of ``Jumping'' of Ext group. This is the ``bottom up'' approach for extension bundles.}
to the direct computation of $H^1({\cal Q})$ and is much easier computationally \cite{Anderson:2011ty}. 
We will illustrate this method with a simple extension bundle in the following subsection.

\subsection{A Rank Two Extension Bundle} \label{sec21}

We assume that the Calabi-Yau threefold $X$ is defined as a hypersurface in a product of projective spaces ${\cal A}=\mathbb{P}^{n_1}\times \ldots \times\mathbb{P}^{n_m}$. We will further ask that it is ``favorable", i.e. the Picard group of $X$ is spanned by the restriction to $X$ of ambient divisors, $D_i$, associated with the hyperplane class in $\mathbb{P}^{n_{i}}$. The defining polynomial of $X$ is given by $p_{0} \in H^0({\cal A}, N)$. Here $N$ is the normal bundle of the hypersurface. Notice that the defining polynomial $p_{0}$ is a redundant description of the complex structure. The independent ones can be obtained from the computation of $H^1(X,TX)$, which will be discussed in section \ref{sec3}.

Consider a rank two bundle $V$ defined as an extension of a line bundle $L$ and its dual
\begin{equation} \label{extB}
 0 \to L \to V \to L^{\vee} \to 0,
\end{equation}
with extension class $\phi \in Ext^1(L^{\vee},L)=H^1(L^2)$. 
If $\phi \neq 0$, the extension is non-trivial and the bundle defined is a an $SU(2)$ indecomposable bundle, otherwise, the extension is trivial and the bundle splits into $V= L \oplus L^{\vee}$. 
However, non-trivial extension does not always exist on $X$ because the Ext group $H^1(L^2)$ that the extension class takes value in depends on the choice of complex structure. We will be interested in a situation where $H^1(L^2)$ is non-trivial only for some special choices of complex structure of $X$ and $H^1(L^2)=0$ for others. Thus, the indecomposable $SU(2)$ bundle is well-defined on special locus in the moduli space and it splits when the complex structure is deformed away from this locus. This setting will play a key role in the moduli stabilization discussed later.

Let's assume that the initial complex structure is chosen such that the $H^1(L^2)$ jumps in dimension and the bundle defined in (\ref{extB}) is a indecomposable $SU(2)$ bundle. At the same time, one needs to choose the initial K\"ahler form $\omega$ of $X$ so that $V$ is poly-stable. Although poly-stability is in general difficult to check, for this simple extension bundle, it can be determined relatively easier \footnote{It have been shown in \cite{Anderson:2009nt}, it is a sufficient condition for $V$ being poly-stable.} by the following criterion \cite{Anderson:2009nt}. 
$V$ is poly-stable if the K\"ahler form $\omega$ satisfies $\mu(L) < 0$ where $\mu(L)$ is the slope of $L$ defined by
\begin{equation}
    \mu(L)=\frac{1}{rank(V)}\int_X c_1(L)\wedge \omega \wedge \omega.
\end{equation}
With this condition, we will fix the initial K\"ahler moduli of $X$ so that $V$ is poly-stable in this example.

Now, let's consider the complex structure deformations with respect to this background. 
From the equation (\ref{atiyahcalss}), deformations that are consistent with the holomorphic bundle is measured by the kernel of the Atiyah class $\alpha \in H^1(V\otimes V^{\vee} \otimes TX^{\vee})$. For this example, it can be shown \cite{Anderson:2009nt} that $\alpha$ is an element of $H^1(X, L^2\otimes TX^{\vee})$, a subset of $H^1(V\otimes V^{\vee} \otimes TX^{\vee})$. 
Thus, the part of sequence in (\ref{H1Q}) containing $\alpha$ becomes 
\begin{equation} \label{AclassSec2}
     \alpha: H^1(X,TX) \to H^2(X, L^2).
\end{equation}
By Bott-Borel-Weil Theorem \cite{hartshorne}, one can express the source $H^1(X,TX)$ and target $H^2(X, L^2)$ in terms of homogeneous polynomials. The kernel of the map $\alpha$ follows by a direct computation of these polynomials. 
However, non-trivial extension does not always exist on $X$ because the Ext group $H^1(L^2)$ that the extension class takes value in depends on the choice of complex structure. We will be interested in a situation where $H^1(L^2)$ is non-trivial only for some special choices of complex structure of $X$ and $H^1(L^2)=0$ for others. Thus, the indecomposable $SU(2)$ bundle is well-defined on special locus in the moduli space and it splits when the complex structure is deformed away from this locus. This setting will play a key role in the moduli stabilization discussed later. 

The idea of ``bottom up'' approach \cite{Anderson:2009nt} is that instead of the direct calculation of ${\rm Ker}(\alpha)$ outlined above, the same information can be obtained by studying the ``jumping'' of Ext group $H^1(X,L^2)$. 
Specifically, recall that the indecomposable bundle $V$ considered here is defined on the special locus in the moduli space where Ext group $H^1(X,L^2)$ jumps, while for generic complex structure, the Ext group $H^1(X,L^2)$ is trivial and $V$ splits. Consider the deformation of complex structure $p_0 \to p_0 + \delta p$. If $\delta P$ is within the special locus of moduli space, the $V$ is still indecomposable and holomorphic. However, if $\delta P$ is generic, the Ext group becomes $H^1(X,L^2) = 0$. It seems that in this case the only choice for extension class is $\phi=0$ and the bundle becomes a direct sum $L \oplus L^{\vee}$. 
In fact, the extension class $\phi \in {H^{1}}(X, L^{2})$, does not change when we vary the complex structure moduli.
Instead, $\phi$ is simply no longer a closed $(0,1)$-form with respect to the new complex structure. 
It means that the extension bundle $V$ after generic deformation $\delta p$ is not holomorphic anymore. In this example, we find that deformations of complex structure that has ``jumping'' Ext group are the same deformations that make $V$ holomorphic. While deformations of complex structure that has trivial Ext group corresponds to ones that make $V$ non-holomorphic. Thus, we see clearly that the calculation of ``jumping'' Ext is equivalent to the Atiyah computation in this example.

Next, let's briefly describe how to identify the special locus where the Ext group $H^1(X, L^2)$ jumps in moduli space. 
By the Koszul sequence \cite{hartshorne}, the line bundle $L^2$ on $X$ can be expressed in terms of bundles on the ambient space ${\cal A}$. 
For the manifold considered here, it is 
\begin{equation}\label{koszul}
    0 \to N^{\vee} \otimes L_{{\cal A}}^2 \stackrel{p_{0}}{\rightarrow} L_{{\cal
    A}}^2 \to L^2 \to 0 \ ,
\end{equation}
where $N^{\vee}$ is the dual of the normal bundle of $X$. 
The cohomology $H^{1}(X, L^2)$ can be computed via the associated long exact sequence
\begin{equation} \label{koszul_long}
0 \to H^1(X, L^2) \to
H^2({\cal A}, N^{\vee}\otimes L_{{\cal A}}^2)
\stackrel{p_{0}}{\rightarrow} H^2({\cal A}, L_{{\cal A}}^2) \to
H^2(X, L^2) \to 0 \ .
\end{equation}
Here $p_0$ is the defining polynomial of $X$. Thus, we have $H^{1}(X, L^2) = \textnormal{Ker}(p_0)$ and $H^{1}(X, {L^{\vee}}^{2}) = \textnormal{Coker}(p_0)$. 
For generic defining polynomial, both $\textnormal{Ker}(p_0)$ and $\textnormal{Coker}(p_0)$ are trivial. 
While for the special choice of $p_0$, they can ``jump" together so that the index $\textnormal{Ind}(L^2)=-h^1(X, L^2)+h^2(X, L^2)$ is preserved \cite{Anderson:2011ty}. 
For the variation of complex structure $p \to p_0 +\delta p$, the number of independent variations that make $h^1(X, L^2)$ jumps is the same as the dimension of ${\rm Ker}(\alpha)$ computed from the equation (\ref{AclassSec2}). 
While the other variations of complex structure with $h^1(X, L^2)=0$ corresponds to ${\rm Im}(\alpha)$ in the Atiyah computation.

This moduli stabilization mechanism can also be understood in the low-energy effective theory. 
Let $\mathfrak{z}_0$ be the initial complex structure moduli chosen within the special locus of moduli space where Ext group of the bundle defined in (\ref{extB}) jumps. 
The non-trivial extension defines a $SU(2)$ bundle. The low-energy gauge group is $E_7$. 
However, if the extension class is trivial, $V$ splits and the low energy the low-energy gauge group is enhanced by an anomalous $U(1)$ factor to $E_7 \times U(1)$ \cite{Sharpe:1998zu,Anderson:2009nt,Anderson:2009sw}. 
The bundle moduli (and matter fields) are charged under the enhanced $U(1)$ symmetry denoted by 
\begin{equation}
 C^i_+ \in H^1(L^{2}),\qquad C^j_- \in H^1({L^{\vee}}^{2}),
\end{equation}
with the subscript $\pm$ indicating the $U(1)$ charge. We will take the extension class $\left<C^i_+ \right> \neq 0$ not far from the zero in $H^{1}(X, L^2)$ and keep $\left<C^j_- \right>=0$ in the following.

The nontrivial F-term comes from the Gukov-Vafa-Witten superpotential \cite{Gukov:1999ya,Anderson:2020ebu}. 
For bundle considered in (\ref{extB}), one can show that superpotential from the contribution of bundle moduli is \cite{Anderson:2011ty}
\begin{equation} \label{superP}
W= \lambda_{ij}(\mathfrak{z}_0) C_+^i C_-^j + \dots  
\end{equation} 
where the coefficients $\lambda_{ij}$ depends on the complex structure moduli $\mathfrak{z}_0^I \in H^1(X,TX)$ explicitly. 
Note that for generic complex structure, the holomorphic function $ \lambda_{ij}(\mathfrak{z}_0)\neq 0$. 
As a result, the fields $C^i_+$ and $C^j_-$ are massive. 
However, on the special locus in complex structure moduli space, $ \lambda_{ij}(\mathfrak{z}_0)=0$, which indicates that $C^i_+$ and $C^j_-$ are massless. The non-trivial vevs of them define the non-Abelian $SU(2)$ bundle \cite{Anderson:2009nt}. This corresponds the ``jump'' of the Ext group described above.

The F-terms of the bundle moduli follows directly from the superpotential above. In particular, the F-terms of $C_+^i$ vanish while the other one
\begin{equation}
 F_{C_-}=\frac{\partial W}{\partial C_- ^j} = \lambda_{ij}(\mathfrak{z}_0) \left\langle C_+^i  \right\rangle \;.
\end{equation}
As discussed above, since the initial complex structure is chosen in the special locus of the moduli space, hence,  $\lambda_{ij}(\mathfrak{z}_0) =0$. 
Consider the fluctuation $\mathfrak{z}_0^{I} \to \mathfrak{z}_0^{I}+\delta \mathfrak{z}$. 
For the fluctuations within the sub-locus where $H^1(X, L^{2})\neq 0$ (denoted by $\mathfrak{z}^{a}_{\parallel}$), the new complex structure is still in the sub-locus and hence $\lambda_{ij}(\mathfrak{z}_0) =0$ and $F_{C_-}$ vanishes. 
However, for the fluctuations away from this sub-locus (denoted by $\mathfrak{z}^{a}_{\perp}$), the new complex structure is a generic element in the complex structure moduli space and $\lambda_{ij}(\mathfrak{z}_0) \neq 0$.
Up to the quadratic order in the field fluctuations, we have the non-zero F-term of $F_{C_-}$ and the potential is 
\begin{equation} \label{box4}
  V = |\frac{\partial
    \lambda_{ij}(\mathfrak{z}_0)}{\partial {\mathfrak{z}^{I}_{\perp}}} \langle C^i_+ \rangle|^{2}|\delta \mathfrak{z}_\perp^{I}|^{2} + \dots \ .
\end{equation}
This gives the mass of the moduli $\mathfrak{z}^{a}_{\perp}$ and these degree of freedoms are stabilized. 
The number of these moduli is the same as the $\textnormal{Im}(\alpha)$ computed in the Atiyah sequence.

In summary, we review the complex structure moduli stabilization in Calabi-Yau compactification due to the presence of holomorphic bundles. The ``true'' moduli are those keeping the holomorphy of the bundle, measured by the kernel of Atiyah class $\alpha$, while the other moduli are stabilized. 
With a simple rank two extension bundle (\ref{extB}), we demonstrate that $\textnormal{Ker}(\alpha)$ can be calculated by an analysis of ``jumping'' Ext group, which is equivalent to the direct computation from the sequence (\ref{H1Q}) and much easier computationally. 
Next, we will focus on the moduli stabilization on an elliptically fibered threefold and study how this mechanism works in that context.

\section{Complex Structure Moduli Stabilization Mechanisms} \label{sec3}

An elliptically fibered Calabi--Yau threefold $X$ consists of a base $B$,
which is a complex two-surface, and an analytic map
\begin{equation}
 \pi:X \to B
 \label{eq:add1}
\end{equation}
with the property that for a generic point $b \in B$, the fiber $E_{b} = \pi^{-1}(b)$ is an elliptic curve. 
In addition, we will require that there exist a global section $\sigma:B \to X$
that assigns to every point $b \in B$ the zero element $\sigma(b)= p \in E_{b}$ discussed below. 
Let $K_{B}$ be the canonical bundle of $B$. 
The elliptic fibration is described by the Weierstrass form
\begin{eqnarray} \label{WeierF}
y^2 - x^3 - f x z^4 - g z^6 = 0,
\end{eqnarray}
where $f$ and $g$ are sections of $K_{B}^{-4}$ and $K_{B}^{-6}$, and $(x,y,z)$ are global sections of $\mathcal{O}_X(2\sigma)\otimes K_{B}^{-2}$, $\mathcal{O}_X(3\sigma)\otimes K_{B}^{-3} $ , $\mathcal{O}_X(\sigma)$ respectively.

Consider a $SU(n)$ vector bundle $V$ on $X$. The Chern character of $V$ can be written as the following general form,
\begin{eqnarray}
ch(V) = n + \sigma \pi^*\eta + \omega [f]+\frac{1}{2}c_3(V),
\end{eqnarray}
where $n$ is the rank of the bundle, $\sigma$ is the section of the fibration, $\eta \in H^2(B, {\mathbb Z})$ is a divisor in the base, $\omega$ in an element in $H^{2,2}(B, {\mathbb Z})$, and $[f]$ is the fiber class of the elliptic fibration. $V$ can be constructed by spectral cover method \cite{Friedman:1997yq,Friedman:1997ih}. It consists of a $n$ sheeted cover of $B$ called
spectral cover ${\cal S}_n$. It is given by finite morphism $\pi_{S}: {\cal S}_n \to B$. 
The class of ${\cal S}_{n}$  can then be written as 
\begin{eqnarray}
[{\cal S}_{n}] = n \sigma + \pi^*(\eta) \in H^2({Z_3}, {\mathbb Z}).
\end{eqnarray}
In affine
coordinates, where $z=1$, we write
\begin{equation}
  \mathcal{S}_n = a_0 + a_2 x + a_3 y + a_4 x^2 + a_5 x^2y + \dots + a_n x^{n/2}=0
\label{speceq}
\end{equation}
If $n$ is odd the last term is $a_nx^{(n-3)/2}y$.
In addition, we need to specify a line bundle $N \to {\cal S}_n$ over the spectral cover called spectral line bundle (or a rank one coherent sheaf generally), and it must satisfy some certain topological conditions. The set of data $(\cal S, N)$ is called spectral data. It is shown \cite{Friedman:1997yq,Friedman:1997ih} that there is a one-to-one correspondence between holomorphic vector bundles and the spectral data,
\begin{eqnarray}
V \leftrightarrow (\mathcal{S}_n, N).
\end{eqnarray}

For an elliptically fibered Calabi-Yau threefold, the complex structure moduli space contains the moduli from the base $B$ and the moduli from the fibration. In fact, one can show that the moduli space of the base $H^1(B,TB)$ is a subset of $H^1(X,TX)$. The detail of the derivation can be found in Appendix \ref{AppCS}.
The complex structure moduli space of the base $H^1(B,TB)$ can be computed from the Euler sequence and adjunction formula. 
Suppose $B$ is defined as a hypersurface of some ambient space $A$, where $A$ can be the direct product of projective spaces $\mathbb{P}^{m_1} \times \ldots \times \mathbb{P}^{m_n}$ or in general toric varieties. 
The defining polynomial of $B$ is $p_0 \in H^0({A}, N)$ for some ample line bundle $N$. 
Then the tangent bundle is described by  
\begin{align} \label{TBF}
     0 \to O_{B}^{\oplus n}  \stackrel{l_1}{\rightarrow} & \bigoplus_{i=1,\dots,n} O_{B}(D_i)^{\oplus(n_i +1)} \to T{A}|_{B} \to 0 \\ \nonumber
    & 0 \to TB \to T{A}|_{B} \stackrel{l_2}{\rightarrow} N \to 0  \nonumber
\end{align}
where $D_i$ are the restriction of the hyperplane divisors of each projective factor of the ambient space $A$ and 
the polynomial maps $l_{1},l_{2}$ satisfy $l_2 \circ l_1 =p_0$. 
The complex structure moduli space $H^1(B, TB)$ follows from the associated long exact sequence. 
From the long exact sequence of cohomology groups, we have 
\begin{equation*}
    0 \to H^0(B,TB) \to H^0(B,TA) \stackrel{l_2}{\rightarrow} H^0(B,N) \to H^1(B,TB) \to H^1(B,TA) \stackrel{l_2}{\rightarrow} H^1(B,N) \to \ldots
\end{equation*}
In general, it contains the polynomial deformations and the non-polynomial ones. 
We will only focus on the polynomial part $H^1_{\textnormal{poly}}(B, TB) \subset H^1(B, TB)$ which is given by
\begin{equation} \label{csM}
    H^1_{\textnormal{poly}}(B, TB)=\frac{H^0(B, N)}{\textnormal{Im}(l_2)} \ .
\end{equation}
The fibration $F$ used in this work has the Weierstrass form. Its complex structure moduli can be computed in the similar manner.

We wish to use these spectral data to find new examples for the complex structure moduli stabilization. There are three types of examples that we are going to consider. In the first type some complex structure deformations are obstructed simply because the spectral cover cannot be defined globally everywhere in moduli space. In the second type, however, the Picard group of the base space jumps over some loci in the complex structure moduli, and if the spectral data depends explicitly on these new divisors, the deformations normal to this loci is obstructed. In the third type, the Picard group of the Calabi-Yau threefold jumps. More clearly, some specific algebraic curves inside the ambient four-fold will lie inside the Calabi-Yau threefold. For some specific bundles, these algebraic curves lie inside the spectral cover.\footnote{Notation: From now on, we will denote the divisors in the base without writing the pullback $\pi^*$ on front, e.g. we will write $\eta$ instead of $\pi^* \eta$. We also denote the spectral cover simply by $[\cal S]$ or $\cal S$ without emphasizing on its degree.}

\subsection{Type I: ``Jumping'' Extension Group} \label{typeI}

Consider the rank two extension bundle $V$ defined in (\ref{extB}). As discussed in Section \ref{sec2}, on special loci in complex structure moduli space, cohomology $H^{2}(X, V\otimes V^{\vee})$ jumps in dimension and the rank of Atiyah class $\alpha$ in (\ref{H1Q}) jumps as well. Hence, complex structure deformations perpendicular to this special loci will make $V$ non-holomorphic and the associated moduli are stabilized \cite{Anderson:2011cza}. In this subsection we consider a similar set-up over an elliptically fibered Calabi-Yau threefold $\pi : X\rightarrow B$, and by analyzing such bundles using spectral cover/Fourier-Mukai construction, one gets a better insight about how exactly such bundles become non-holomorphic. 

As in \cite{Anderson:2019agu} we take $\mathcal{L}=\mathcal{O}_{X}(-\sigma+D_b)$. The extension group of $V$ is given by 
\begin{align}\label{extv2}
  Ext_{X}^1(\mathcal{L}^{\vee},\mathcal{L})&=H^1(X,\mathcal{O}_{X}(-2\sigma +2D_b)) \\ \nonumber
 &\simeq H^0(B,\mathcal{O}_{B}(2D_b+K_B))\oplus H^0(B, \mathcal{O}_{B}(2D_b-K_b)),  
\end{align}
where the last isomorphism is derived by the Leray spectral sequence relative to the morphism $\pi$. It is also not hard to show \cite{Anderson:2011cza} that the (holomorphic) deformations of $V$ i.e., $H^{1}(X, V^{\vee}\otimes V)\simeq H^2(X, V^{\vee}\otimes V)$ depends on the extension group above.

On the other hand, over the elliptically fibered Calabi-Yau $X$ the vector bundle $V$ can be parameterized by the spectral data $(\mathcal{S},N)$ \cite{Friedman:1997yq,Friedman:1997ih}, where $\cal S$ is a double cover of the base $B$ and $N$ is a rank one coherent sheaf over $\cal S$. When $\cal S$ is smooth, $N$ is simply a line bundle. The divisor class of $\cal S$ is determined by the topology of $V$, 
\begin{align}
[\cal S]&= 2\sigma + \eta, \\
\eta&= 2D_b + c_1(B).  
\end{align}
Also the general form of the algebraic equation of the spectral cover (when $X$ is given by the Weierstrass fibration) is given by, 
\begin{align}\label{specV2}
&\mathcal{S} = a_2 Z^2+ a_0 X, \\ 
&a_0\in H^0(B,\mathcal{O}_{B}(c_1(B)+2D_b)),\quad a_2 \in H^0(B,\mathcal{O}_{B}(-c_1(B)+2D_b)).  
\end{align}
By computing the Fourier-Mukai transformation of $V$, it is shown \cite{Anderson:2019agu} that the polynomials $a_2$ and $a_0$ can be identified with the elements of the extension group \eref{extv2}. Therefore in the spectral cover language the loci in the complex structure moduli where the group $H^{2}(X, V^{\vee}\otimes V)$ jumps correspond to the loci where it is possible to add new terms in the polynomials $a_2$ and $a_0$. Once one add such terms in the spectral cover the complex structure moduli is stabilized. Since the the complex structure deformations that turns off such terms make the spectral cover non-algebraic and hence the associated vector bundle, non-holomorphic. 

\subsubsection*{Example}

To be more concrete consider a Weierstrass elliptically fibered Calabi-Yau threefold $X$ over the following surface, 

\begin{eqnarray} \label{s3base}
 B=\left[
 \begin{array}{c|ccc}
 \mathbb{P}^{1}_{x}&2\\
 \mathbb{P}^{2}_{y}&2
 \end{array}
 \right], \quad c_1(B)= H_2.
\end{eqnarray}
Let us choose $D_b=-H_1+2H_2$, where $H_1$ and $H_2$ are hyperplanes in $\mathbb{P}^{1}_{x}$ and $\mathbb{P}^{2}_{y}$ respectively. Therefore 
\begin{align}
a_0&\in H^0(B,\mathcal{O}_{B}(-2,5))\;, \\
a_2&\in H^0(B,\mathcal{O}_{B}(-2,3))\;. 
\end{align}

Generic defining equation $P$ for $B$ can be written as
\begin{eqnarray}\label{Ptype1}
P=x_1^2P_1(y)+x_2^2P_2(y)+ x_1x_2 P_3(y),
\end{eqnarray}
where $P_1$, $P_2$, $P_3$ are degree two homogeneous polynomials in y. However the cohomology of line bundles over $B$ can jump over codimension one loci in complex structure moduli of the base $B$. In particular the cohomology of $\mathcal{O}_{B}(-2,3)$ and $\mathcal{O}_{B}(-2,5)$ can jump as,\footnote{The surface $B$ is still smooth over this special complex structure locus.} 
\begin{eqnarray}\label{h0jump}
&    h^0(B,O(-2,3)) =
\begin{cases}
0, & P\quad \text{generic} \\
3, & P=x_1^2P_1(y)+x_2^2P_2(y)
\end{cases},   &\\ \nonumber
&h^0(B,O(-2,5)) =
\begin{cases}
9, & P\quad \text{generic} \\
13, & P=x_1^2P_1(y)+x_2^2P_2(y)
\end{cases}.&
\end{eqnarray}

The computation details are in the Appendix \ref{AppTypeI}. 
This means the spectral cover in general can be written as,
\begin{eqnarray}
\cal S=\begin{cases}
a_0 Z^2, & P\quad \text{generic} \\
a_0 Z^2 + a_2 X, & P=x_1^2P_1(y)+x_2^2P_2(y),
\end{cases}
\end{eqnarray}
where in the second case $a_0$ is a rational polynomial \cite{Anderson:2015iia} which is holomorphic over the hypersurface $P=0$,
\begin{eqnarray}\label{newa0}
a_2 = \frac{P_1(y)}{x_2^2} y_1, \frac{P_1(y)}{x_2^2} y_2, \frac{P_1(y)}{x_2^2} y_3.
\end{eqnarray}
Similarly, the extra terms for the special defining equation $P=0$ is generally of form $\frac{P_1(y)}{x_2^2} f_3(y)$. Where $f_3$ is degree three polynomial in $y_1,y_2,y_3$, but we don't need the detailed form of it.  
Therefore in the complex structure locus where $P=x_1^2P_1(y)+x_2^2P_2(y)$ the extension group \eref{extv2} jumps, and it is possible turn on these new terms. This corresponds to add the new terms in $a_0$ and $a_2$ in the spectral cover. We consider the new terms in $a_0$ and $a_2$ in turn. 

First consider the $a_2$ term. When we turn on terms \eref{newa0}, the spectral cover is smooth. As it is shown in \cite{Friedman:1997yq,Friedman:1997ih,Anderson:2019agu} generally the spectral line bundle $N$ is
\begin{eqnarray}
N=\mathcal{O}_S(\sigma+D_b).
\end{eqnarray}
Therefore in general $N$ depends on the divisor $\sigma|_S$. The image of this curve on $B$ is the curve $(a_2=0)$. When one turns on the term $x_1x_2P_3(y)$ in \eref{P} This curve ``disappears". In other words, away form the locus (in the complex structure moduli space) where $P_3(y)$ in \eref{P} is nonzero the divisor $-2H_1+3H_2$ is not effective anymore. This means after this complex structure deformation the curve $(a_2=0)$ mixes with $(2,0)$ or $(0,2)$ cycles. Therefore the corresponding curve $\sigma|_S$ on the spectral cover is not algebraic anymore and mixes with $(2,0)$ or $(0,2)$ cycles. Hence the corresponding vector bundle becomes non-holomorphic. 

Next, suppose $a_2=0$. Instead we turn on the extra terms in $a_0$. In this case the spectral cover\footnote{In this case the spectral cover is a union of two copies of the zero section $Z=0$ (which are ``infinitesimally close" to each other) and the vertical surface $a_2=0$.} $S=a_0 Z^2$. Similar to the previous case if $a_0$ depends on the terms $\frac{P_1(y)}{x_2^2} f_3(y)$, the corresponding curve $a_0$ becomes non-algebraic after the complex structure deformation. Therefore the corresponding vector bundle $V$ becomes non-holomorphic after the deformation.
 
So the usual stabilization scenario of complex structure moduli by rank two vector bundles \cite{Anderson:2011cza,Gray:2019tzn} translates to the question of whether the coefficients of the spectral cover equation exists globally. Therefore one may generalize this type of stabilization to vector bundles of arbitrary ranks simply by asking if a holomorphic vector bundle with the Chern character $ch(V)=n+\sigma \eta+ \omega [f]$ is given, then does the divisor $[S]= n \sigma + \eta$ have global section everywhere in the complex structure moduli?

Now, we will see how many complex structure moduli are stabilized in this example. 
if the initial defining polynomial is fixed by equation (\ref{specialCS}), the complex structure moduli space $H^1(B,TB)$ can be computed by equation (\ref{TBF}) and (\ref{csM}). It follows that the space of polynomial deformations $H^1_{\textnormal{poly}}(B,TB)$ has dimension $6$ with basis $\{d_i\},i=1,2,\ldots,6$. 
As discussed above, if any monomials in a polynomial deformation of $P$ contains $x_{0}x_{1}$ term and spectral cover depends on the extra terms in \eref{h0jump}, then $\cal S$ will be non-algebraic after the deformation and $V$ is not holomorphic any more. 
It turns out that there are $4$ such deformations in $\{d_1,d_2,\ldots,d_6\}$.\footnote{Note that by an argument from linear algebra, this result does not change by choice of basis in $H^1_{\textnormal{poly}}(B,TB)$.} 
Therefore, we find $4$ out of $6$ complex structure moduli of $B$ are stabilized.

\subsection{Type II: ``Jumping'' Picard Group }\label{typeII}

For elliptically fibered Calabi-Yau threefold, the Picard group of the base $B$ can jump in dimension, which can be used for moduli stabilization. The idea is that look for the loci in the complex structure moduli that Picard group jumps (Noether-Lefschetz loci) and choose the spectral data to depend on the ``new" elements of the Picard group. Therefore deforming the complex structure back is impossible because the corresponding vector bundle becomes non-holomorphic. We will explain this approach by means of an example. 

\subsubsection*{Example}

More concretely consider the following surface
\begin{eqnarray}\label{iniSurf}
B^0 = \begin{array}{cccccc|c}
x_1 & x_2 & z_1 & z_2 & u_1 & u_2 & P \\
\hline
1 & 1 & 0 & 0 & 0 & 0 & 1 \\
0 & 1 & 1 & 1 & 0 & 0 & 3 \\
0 & 1 & 0 & 1 & 1 & 1 & 3 
\end{array}
\end{eqnarray}
Where P is generically given by
\begin{eqnarray}\label{P}
P=x_1 (z_1^3 f_3(u)+z_1^2 z_2 f_2(u)+ z_1 z_2^2 f_1+z_2^3 f_0)+x_2(z_1^2 g_2(u)+z_1 z_2 g_1(u)+ z_2^2 g_0)=0.
\end{eqnarray}
Here, $f_d(u)$ and $g_d(u)$ represent degree $d$ homogeneous polynomials in $\{u_1,u_2\}$. 
This is actually the blow-up of $\mathbb{F}_1$ at the following six points
\begin{eqnarray}
z_1^3 f_3(u)+z_1^2 z_2 f_2(u)+ z_1 z_2^2 f_1+z_2^3 f_0 =  z_1^2 g_2(u)+z_1 z_2 g_1(u)+ z_2^2 g_0 =0.
\end{eqnarray}
However the Picard number of $B$ is three ($\rho=3$), but $h^{1,1}=8$. In other words this surface is not favorable. This is because the six points are blown up ``at the same time". Note that the exceptional divisor can be identified by $-J_1+2 J_2+2 J_3$ is effective and has self intersection $(-6)$. The global section of the corresponding line bundle is given as
\begin{eqnarray}
e=\frac{z_1^2 g_2(u)+z_1 z_2 g_1(u)+ z_2^2 g_0}{x_1}.
\end{eqnarray}
The zero locus $e=0$ corresponds to a union of six $\mathbb{P}^1$'s each one is a $(-1)$ curve. 

 Now let's tune the complex structure of the base to\footnote{The hypersurface is still smooth after this tuning.}
\begin{eqnarray}\label{tunedP}
P=x_1 (z_1^3 f_3(u)+z_1^2 z_2 f_2(u)+ z_1 z_2^2 f_1+z_2^3 f_0)+x_2(z_1 h_1(u)+z_2 h_0)(z_1 l_1(u)+z_2 l_0)=0.
\end{eqnarray} 
Now there are two independent set of three curves over 
\begin{eqnarray}
z_1^3 f_3(u)+z_1^2 z_2 f_2(u)+ z_1 z_2^2 f_1+z_2^3 f_0 = z_1 h_1(u)+z_2 h_0 =0, \nonumber
\end{eqnarray}
and over 
\begin{eqnarray}
z_1^3 f_3(u)+z_1^2 z_2 f_2(u)+ z_1 z_2^2 f_1+z_2^3 f_0 = z_1 l_1(u)+z_2 l_0 =0.
\end{eqnarray}
So the Picard number jumps to 4. To make this more clear one can rewrite the same geometry in the following way
\begin{eqnarray}
B^1 &=& \begin{array}{cccccccc|cc}
y_1 & y_2 & x_1 & x_2 & z_1 & z_2 & u_1 & u_2 & P_1 & P_2 \\
\hline
1 & 1 & 0 & 0 & 0 & 0 & 0 & 0 & 1 & 1 \\
0 & 0 & 1 & 1 & 0 & 0 & 0 & 0 & 1 & 0 \\
0 & 2 & 0 & 1 & 1 & 1 & 0 & 0 & 2 & 3 \\
0 & 2 & 0 & 1 & 0 & 1 & 1 & 1 & 2 & 3
\end{array}.\\
P_1&=& y_1 x_2 (z_1 h_1(u)+z_2 h_0) + y_2 x_1 =0, \nonumber \\
P_2&=&-y_1(z_1^3 f_3(u)+z_1^2 z_2 f_2(u)+ z_1 z_2^2 f_1+z_2^3 f_0)+ y_2 (z_1 l_1(u)+z_2 l_0) =0.
\end{eqnarray}
The geometry above is equivalent to \eref{tunedP} by eliminating $y_1$ and $y_2$. The two $(-1)$-curves and their global sections can be identified as
\begin{eqnarray}\label{extraE}
&e_1 = -J_1 +J_3+J_4, \qquad &a_1 =\frac{z_1 l_1(u)+z_2 l_0}{y_1}, \\
&e_2 = J_1-J_2+J_3+J_4,\quad &a_2 = \frac{y_1 (z_1 h_1(u)+z_2 h_0)}{x_1}.
\end{eqnarray}
Now, we want to use this jump in $Pic(B)$ to stabilize the complex structure deformation that makes the divisor $e_1+e_2$ irreducible. To do this, consider for example a holomorphic vector bundle with Chern character
\begin{eqnarray} \label{VtypeII}
ch(V)=n-(\sigma (\eta_0 + e_1)+\omega [f]) +\dots,
\end{eqnarray}
where $\eta_0$ depends on $e_1+e_2$. Due to the special form of $\eta=\eta_0+e_1$, the spectral cover of $V$ is reducible. Therefore the spectral sheaf $\cal N$ is therefore defined as the extension,
\begin{eqnarray}
0\rightarrow  {\cal L}_1 \rightarrow  {\cal N} \rightarrow {\cal L}_2 \rightarrow 0,
\end{eqnarray}
where ${\cal L}_1$ is a line bundle supported on $\pi^*e_1$, and ${\cal L}_2$ is a line bundle supported over smooth surface in the divisor class $n\sigma + \eta_0$. After a calculation similar to \cite{Anderson:2019agu} one can show,\footnote{The term $\lambda_1 e_1$ in $c_1(\mathcal{L}_2)$ have to be compensated with a term $-\lambda_1 {\cal S}_0$ in $c_1(\mathcal{L}_1)$. They contribute to $c_3(V)$ and $\omega$, but since these are not important for the main purpose of this paper, we ignore further details.}
\begin{eqnarray}
c_1({\cal L}_2) &=& \frac{1}{2}(c_1(B)-[{\cal S}_0]) + \lambda \left(n\sigma - \eta_0 +n c_1(B) \right) + \lambda_1 e_1 .
\end{eqnarray}
The details to derive it is in Appendix \ref{AppFM}. What is important here is the dependence of the spectral sheaf on $e_1$. This means deforming the complex structure back to the point where $e_1+e_2$ is irreducible is impossible for two reason. First the spectral cover $[{\cal S}] = [{\cal S}_0]+e_1$ becomes non-effective, i.e. one cannot define the spectral cover globally. Second if $\lambda_1\ne 0$ the line bundle ${\cal L}_2$ over ${\cal S}_0$ becomes non-holomorphic as we deform the complex structure.

We will study how the complex structure moduli of $B^0$ are stabilized in this example. 
Set the initial defining polynomial $P^{(0)}$ to be the special polynomial defined in (\ref{tunedP}). 
To simply the notation, let's write it as $P^{(0)} = x_1 F_1^{(0)} + x_2 F_2^{(0)}$ where 
$F_1^{(0)}=z_1^3 f_3(u)+z_1^2 z_2 f_2(u)+ z_1 z_2^2 f_1+z_2^3 f_0$ and $F_2^{(0)}$ is a reducible polynomial given by
\begin{eqnarray}\label{reducibleF}
F_2^{(0)} = \left( A_1 u_1 z_1 + A_2 u_2 z_1 + A_3 z_2 \right) \left( B_1 u_1 z_1 + B_2 u_2 z_1 + B_3 z_2 \right),
\end{eqnarray} 
with $A_i$'s and $B_i$'s the coefficients. As discussed above, the base $B^0$ admits two algebraic divisors $e_1$ and $e_2$ given by equation (\ref{extraE}). With them, one can define a holomorphic bundle $V$.

Now, consider the polynomial deformations of $P^{(0)}$ given by $\delta P = x_1 \delta F_1 + x_2 \delta F_2$. At first, we keep our discussion in general and allow $\delta P$ to be any elements in $ H^0(B^0,N)$. Eventually, we will restrict them to be the deformation in $H^1_{\textnormal{poly}}(B^0,TB^0)$. 
To keep $V$ holomorphic, the term $F_2^{(0)} + \delta F_2$ in deformed defining polynomial $P^{(0)} + \delta P$ should remain reducible. Deformations satisfying this condition are called {\it reducible deformations}. Before studying complex structure moduli stabilization, one needs to find all the reducible deformations.

Obviously, $\delta P$ with $\delta F_2 = 0$ are reducible deformations because they have no effect to $F_2^{(0)}$. 
Let $M^1_{\;\;r}$ be the space of all such deformations. 
For deformations with $\delta F_2 \neq 0$, it is not clear if $F_2^{(0)} + \delta F_2$ remain reducible. However, we know that the deformed $F_2$ term, if reducible, should be able to be written in the form of (\ref{reducibleF}).
The first-order deformation $\delta F_2$ just shifts the initial coefficients in (\ref{reducibleF}) to a set of new coefficients $ A_i+\delta A_i$ and $ B_i+\delta B_i$. We will see this in details in Appendix \ref{appT2}.
Notice that if $\delta F_2$ is a deformations that keeps of $F_2^{(0)}+\delta F_2$ reducible, then $\delta P = x_2\delta F_2$ is a reducible deformation of $P^{(0)}$. Without losing clarity, these $\delta F_2$ will also be called reducible deformations.

To obtain the reducible deformations of $F_2^{(0)}$, it is sufficient to calculate what is the $\delta F_2$ obtained by a shift of each coefficients. 
For example, the reducible deformation corresponding to a shift of the first coefficient $A_1 + \delta A_1$ is obtained by  
\begin{eqnarray*}
    F_2 & = & \left( (A_1+\delta A_1) u_1 z_1 + A_2 u_2 z_1 + A_3 z_2 \right) \left( B_1 u_1 z_1 + B_2 u_2 z_1 + B_3 z_2 \right) \\
    & = & F_2^{(0)} + \delta A_1 u_1 z_1 \left( B_1 u_1 z_1 + B_2 u_2 z_1 + B_3 z_2 \right) \\
    & = & F_2^{(0)} + \delta A_1 e^A_1\;.
\end{eqnarray*}
In the similar manner, one can obtain the reducible deformations from shifting the other coefficients 
\begin{eqnarray} \label{es6}
    &e^A_1  = u_1 z_1 \left( B_1 u_1 z_1 + B_2 u_2 z_1 + B_3 z_2 \right),  
    &e^B_1  = u_1 z_1 \left( A_1 u_1 z_1 + A_2 u_2 z_1 + A_3 z_2 \right)\\ \nonumber
    &e^A_2  = u_2 z_1 \left( B_1 u_1 z_1 + B_2 u_2 z_1 + B_3 z_2 \right),  
    &e^B_2  = u_2 z_1 \left( A_1 u_1 z_1 + A_2 u_2 z_1 + A_3 z_2 \right)\\ \nonumber
    &e^A_3 = z_2 \left( B_1 u_1 z_1 + B_2 u_2 z_1 + B_3 z_2 \right), \quad 
    &e^B_3  = z_2 \left( A_1 u_1 z_1 + A_2 u_2 z_1 + A_3 z_2 \right) \nonumber
\end{eqnarray}
The linear combination of $\{e^A_i\}$ and $\{e^B_j\}$ give a reducible deformations $\delta F_2$. 
However, there is a relation among them 
\begin{eqnarray} \label{relationR}
    F_2^{(0)}+A_1 e^A_1 + A_2 e^A_2 + A_3 e^A_3 & = & F_2^{(0)} + B_1 e^B_1 + B_2 e^B_2 + B_3 e^B_3 
\end{eqnarray}
as one can easily check with the equation (\ref{reducibleF}) and (\ref{es6}). 
Thus, only $5$ of $\{e^A_i\}$ and $\{e^B_j\}$ are independent and, when multiplying them by $x_2$, they span a vector space denoted by $M^2_{\;\;r}$ containing all the reducible deformations with $\delta F_2 \neq 0$. Therefore, the space of all the reducible deformations for $P^{(0)}$ is $M_r = M^1_{\;\;r} \cup M^2_{\;\;r}$.

The complex structure moduli space $H^1(B^0,TB^0)$ is computed with the equation (\ref{TBF}) and (\ref{csM}). 
It follows that the space of polynomial deformations $H^1_{\textnormal{poly}}(B^0,TB^0)$ has dimension $5$ \footnote{There are also non-polynomials deformations, but we will not consider them here.}, spanned by homogeneous polynomials $d_i,i=1,2,\ldots,5$. The intersection $H^1_{\textnormal{poly}}(B^0,TB^0) \cap M_r$ contains all the polynomial deformations from $B^0$ that keep $V$ holomorphic and contribute to the nontrivial kernel of Atiyah class in (\ref{H1Q}). The dimension of this intersection determines the number of the "true" moduli in the effective theory while the moduli in the complement of the intersection are stabilized. In our example, $M_r \cap H^1_{\textnormal{poly}}(B^0,TB^0)$ is a $4$ dimensional space. Thus, $1$ complex structure moduli is stabilized.\footnote{One could guess this result by comparing the complex structure of $B^0$ and $B^1$. Since the complex structure of $B^1$ is ``frozen" in the sense that it is impossible to ``glue" the divisors $e_1$ and $e_2$ into a single divisor $e$, but any other complex structure deformation of $B^0$ is in a one-one correspondence with the complex structure deformations of $B^1$. So the space $H^1(TB^1)$ actually corresponds to the flat (unstabilized) directions. Finally one can check that $h^1_{poly}(B^1,TB^1)-h^1_{poly}(B^0,TB^0)=-1$ as expected.}

\subsection{Type III: Emergent Algebraic Cycle} \label{typeIII}

In the final example we let an algebraic curve to exist in total elliptic fibration after tuning the complex structure of the Calabi-Yau threefold. As in the previous example, the complex structure can be stabilized by choosing a vector bundle such that it's topology depends on this new algebraic curve. 

Consider an elliptic fibered Calabi-Yau threefold $\pi: X \rightarrow B$ with Weierstrass fibration given by
\begin{eqnarray}\label{Fdef}
F^{(0)}_W = y^2-x^3-f x z^4-g z^6 = 0\;, \quad g = \alpha g' \;,
\end{eqnarray}
where $f \in H^0(B,K_{B}^{-4})$ and $g \in H^0(B,K_{B}^{-6})$ chosen to be a product of two polynomials $\alpha$ and $g'$ with correct degree. This reducible polynomial $g$ allow us to define an algebraic 2-cycle on $X$ 
\begin{eqnarray} \label{AcycleIII}
[C] = \lbrace x=y=\alpha=0 \rbrace.
\end{eqnarray}
Note that this algebraic cycle is independent of the intersection of divisors $\sigma \cdot \pi^* D_b$ and the fiber class $[f]$. Now consider the spectral cover of a general rank $n$ vector bundle $V_n$
\begin{eqnarray}
\mathcal{S}(V_n)=a_0 z^n + a_{2} z^{n-2}x+ \dots + a_n x^{n/2}.
\end{eqnarray}
The 2-cycle $[C]$ intersect with spectral cover at finite number of points. However after the choosing $a_0$ to be reducible
\begin{eqnarray}\label{Sdef}
    a_0 =\alpha a_0,
\end{eqnarray}
the curve $[C]$ can lie inside the spectral cover. So if the spectral sheaf $N$ depends on this new divisor inside $\mathcal{S}(V_n)$ we can stabilize the complex structure moduli, simply because any deformation that removes $[C]$ from $\mathcal{S}(V_n)$ makes $N$ and hence the bundle $V_n$ non-holomorphic. However a Fourier-Mukai analysis as in \cite{Anderson:2019agu} shows that if the Chern classes of the vector bundle does not depend on the 2-cycle $[C]$, then the spectral line bundle also does not depend on the new divisor $[C]$ in $\mathcal{S}(V_n)$, and therefore the complex structure moduli cannot be stabilized. So suppose the Chern character of $V_n$ is of the form (See Appendix \ref{AppFM} for details)
\begin{eqnarray} \label{ChernChIII}
Ch(V_n)=n-\left(\sigma \cdot \eta+\omega [f]+[C]\right)+\frac{1}{2}c_3(V_n).
\end{eqnarray}
By doing a similar calculations as in \cite{Anderson:2019agu} we can derive the topological constraints on the divisor class of $S(V_n)$, Chern classes of $V_n$ and $\mathcal{L}$. We will not repeat them here, however we just mention the part of the results that we will need,\footnote{Note that we should choose $\alpha$ such that $\frac{1}{n} \pi_*[C]$ be integral.}
\begin{eqnarray} 
[\mathcal{S}(V_n)] &=& n\sigma + \eta + \pi_*[C], \label{specC}\\
c_1(\mathcal{N}) &=& \frac{1}{2}\left([S(V_n)]-c_1(B)\right) + \lambda \left(n\sigma-\eta+n c_1(B)\right) +\left(\frac{1}{n}\pi_*[C] -[C] \right).\label{lineC}
\end{eqnarray}
Therefore $\mathcal{N}$ explicitly depends on the divisor $[C]$ in the spectral cover. This dependence (partially) stabilize both complex structure and vector bundle moduli.

Consider a deformation of Weierstrass fibration $F^{(0)}_W \to F^{(0)}_W + \delta g z^6$, which induces a deformation to function $g$ by 
\begin{eqnarray}
\alpha g'&\longrightarrow& g.
\end{eqnarray}
If the deformed $g$ can be expressed as a product of two polynomials (having the same degree with $\alpha$ and $g'$), then the algebraic curve $[C]$ defined in (\ref{AcycleIII}) still exist and the bundle $V$ remain holomorphic. However, if the function $g$ after deformation is irreducible, the curve $[C]$ is not algebraic anymore, and the vector bundle defined by the spectral line bundle \eref{lineC} which explicitly depends on $[C]$ becomes non-holomorphic. Therefore the associated complex structure moduli of $F_W$ can be stabilized.  
As we will explain in Section \ref{sec4}, these polynomial deformations maps to complex structure moduli in the F-theory dual picture, and they are stabilized by a four-flux.

\subsubsection*{Example}

Let $\pi: X \rightarrow \IP^2$ be an elliptically fibered Calabi-Yau threefold with base $\IP^2$. The fibration is Weierstrass introduced in equation (\ref{WeierF}) with $f \in H^0(\IP^2,O(12))$ and $g \in H^0(\IP^2,O(18))$.
Let $\{u_1,u_2,u_3\}$ be the homogeneous coordinates of $\IP^2$. Choose the initial fibration to be 
The initial fibration is chosen to be
\begin{equation} \label{FIII}
    f^{(0)} = F_{12}(u),\quad g^{(0)} = \alpha F_{16}(u)
\end{equation}
where $F_{12}$ and $F_{16}$ are generic polynomials with degree $12$ and $16$ and $\alpha$ is a degree $2$ homogeneous polynomial. 
As shown in (\ref{AcycleIII}), the fibration with reducible $g$ admits an algebraic curve $[C]$ in $\IP^2$, which will be used to define the topology of vector bundles.

Consider a $SU(2)$ extension bundle $V$ over $X$. The spectral cover construction of $V$ introduced in equation (\ref{specV2}) is $S= a_2 z^2+ a_0 x$ where $x,z$ are coordinates of the fibration. If $D_b=O(2)$, then $a_2$ and $a_0$ should be global sections of $O_B(7)$ and $O_B(1)$. Let's take them to be 
\begin{align}
    a_0 &=  F_1(u_1,u_2,u_3),\quad a_2 = \alpha F_5(u_1,u_2,u_3)
\end{align}
where $F_d$ is a generic homogeneous polynomial with degree $d$. 
As discussed in (\ref{ChernChIII}) and (\ref{lineC}), if the Chern character of $V$ depends on the algebraic 2-cycle $[C]$, then the Chern class of spectral sheaf depends on $[C]$ explicitly.  
Thus, the holomorphy of $V$ depends on the existence of the algebraic cycle $[C]$ that will be used to stabilize complex structure moduli in a second.

We will study how many complex structure moduli can be stabilized in this example. The clliptically fibered Calabi-Yau manifold $X$ can be regarded as 
\begin{eqnarray}\label{ECYex3}
X = \begin{array}{cccccc|c}
u_1 & u_2 & u_3 & x & y & z & P \\
\hline
1 & 1 & 1 & 6 & 9 & 0 & 18 \\
0 & 0 & 0 & 2 & 3 & 1 & 6 
\end{array}
\end{eqnarray}
With it, one can compute the complex structure moduli space $H^1(X,TX)$ from the equation (\ref{TBF}) and (\ref{csM}). Again, we will only focus on the polynomial deformations $H^1_{\textnormal{poly}}(X,TX)$ which has dimension $272$. 
Since the base $\IP^2$ is rigid, all the complex structure moduli are coming from the fibration, i.e. deformations of $f$ and $g$ \footnote{Indeed, the degree of freedom of $f$ and $g$ is $h^0(\IP^2,O(12))+h^0(\IP^2,O(18))=91+190=281$. Considering the coordinates redefinition of $\IP^2$, one find the independent degree of freedom of $f$ and $g$ is $272$, which matches the number of complex structure moduli counted in $h^1(X,TX)=272$.}.


As discussed above, deformations that keep $V$ holomorphic are those preserve the reducibility of $g= \alpha F_{16}(u)$. Again, we will call them {\it reducible deformations}. Obviously, any polynomial without $z^6$ terms is reducible deformation. While, the reducible deformations with $z^6$ terms can be calculated from the method in Appendix \ref{appT2}. We find that these deformations can be generated from $159$ polynomial deformations $e^A_i,i=1,2,\dots,6$ and $e^B_j,i=1,2,\dots,153$, which are obtained by deforming the coefficients in $\alpha$ and $F_{16}(u)$ respectively. In fact, the space of these deformations denoted by $M^2_{\;\;r}$ has dimension $158$ because there is a nontrivial relation in (\ref{genRelation}) between $\{e^A_i\}$'s and $\{e^A_j\}$'s.

Finally, we will see how many polynomial deformations $H^1_{\textnormal{poly}}(X,TX)$ are stabilized. 
One needs to split $272$ deformations found earlier into two groups. The former ones does not terms proportional to $z^6$ which have no effect on $g$ and should be considered as the moduli of the effective theory. There are $82$ of them. The latter ones contain terms proportional to $z^6$ and could have an effect on the reducibility of $g$. It turns out that there are $190$ such polynomials. Their coefficients of $z^6$ is denoted by $d_i,i=1,2,\ldots,190$. 
Then, one needs to study how many of them or their linear combinations can give rise to reducibility preserving deformations in $M^2_{\;\;r}$. 
Direct computations shows that all $158$ polynomials in $M^2_{\;\;r}$ can be expressed in the complex structure moduli $\{d_i\}$.
Thus, there are $241$ polynomial deformations that preserves the reducible of $g$ while there are $32$ complex structure moduli are stabilized.

\section{Discussion and Comments on F-Theory Dual} \label{sec4}
In this section we briefly describe the F-theory translation of the Heterotic examples mentioned before. When spectral cover is smooth the duality procedure is well known \cite{Donagi:2008ca}. The geometry of the F-theory half-Calabi-Yau fourfold (more precisely $dP_8$ fibration) is given by (for a $SU(n)$ bundle with $n\le 5$), 
\begin{eqnarray}\label{FthGeo}
Y^2=X^3+f X v^4+g v^6 + u v^{5-n} S_n (X,Y,v),
\end{eqnarray}
where the term in the parenthesis is the spectral cover of a $SU(n)$ vector bundle (with $z$ is replaced by $v$), and $(X,Y,u,v)$ are the homogeneous coordinates of $\mathbb{WP}^{3211}$. In addition using the cylinder map one constructs the four-flux. The details can be found in \cite{Donagi:2008ca}. It is only sufficient mention that the part of the $c_1(\mathcal{L})$ that lies in the kernel of $\pi_S : S \rightarrow B$, lifts to the four-flux in F-theory language, 
\begin{eqnarray}\label{specdata}
\begin{array}{c}
c_1(\mathcal{L})=\frac{1}{2}\left(c_1(B)-[S]\right) + \gamma \\
\pi_{S*} \gamma= 0
\end{array}\Bigg\} \Rightarrow \gamma \longrightarrow G_4 \sim P_{Cyl} ^* \gamma,
\end{eqnarray}
where $P_{Cyl}$ is the cylinder map which is a $\mathbb{P}^1$ fibration over the spectral cover $S_n$,
\begin{eqnarray}
    P_{Cyl}:\{ {S_n(X,Y,v)=0}\subset X_F\} \xrightarrow{\mathbb{P}^1} S_n(x,y,z) \subset X_{Het}.
\end{eqnarray}
In terms of the local 7-brane data, there is a doublet $(E,\Phi)$, called Higgs bundles, which consists of a holomorphic bundle $E$ on $B$ (where 7-brane wraps) and a ``Higgs" field $\Phi \in H^0(E\otimes E^* \otimes K(\eta))$ which satisfy the Hitchin system equations \cite{Beasley:2008dc}. One can study the Higgs bundles by spectral data,
\begin{eqnarray}
    (E,\Phi) \leftrightarrow (S_H,\mathcal{L}_H),
\end{eqnarray}
where $S_H$ is a finite cover of $B$ defined as subvariety of $Tot(K)$,
\begin{eqnarray}
    S_H=Det\left( \lambda 1-\Phi \right)=0,
\end{eqnarray}
where $\lambda$ is the tautological section of $Tot(K)$, and $\mathcal{L}_H$ is a line bundle over $S_H$. The Higgs bundle $(E,\Phi)$ can be reconstructed by spectral data as,
\begin{eqnarray}
  &&  E = \pi_* \mathcal{L}_H,\quad \Phi =\pi_* \lambda, \\
  &&  \pi_S: S_H \rightarrow B.
\end{eqnarray}
The Higgs bundle spectral data is clearly very similar to the Heterotic spectral data derived from the Fourier transform of $V$. They are indeed related to each other by cylinder construction \cite{Donagi:2009ra}. 

Finally note that what is responsible for the complex structure is the F-term derived from the Gukov-Vafa-Witten like superpotentials, which takes the following forms in Heterotic, F-theory and 7-brane
\begin{center}
    \begin{tabular}{ccccc}
    
        Heterotic & $\leftrightarrow$ & F-theory & $\leftrightarrow$ & 7-brane   \\
        & & & &\\
         $W=\int_{X_{Het}} \omega_3\wedge \Omega_3$ & $\leftrightarrow$ & $W=\int_{X_F} G_4\wedge \Omega_4$ &$\leftrightarrow$ & $W= \int_{B} Tr(F_2 \wedge \Phi)$ \\
         
    \end{tabular}
\end{center}
Where $\Omega$ corresponds to the complex structure top-form and $F_2$ is the curvature two-form of $E$. With this introduction, we move forward to explain briefly the ``image" of the moduli stabilization scenarios of Heterotic string in the dual F-theory. Locally over the 7-brane, the story is similar to the Heterotic version. Basically, the deforming away from special loci in complex structure moduli of the base make the Higgs bundle non-holomorphic, and the term $F_2 \wedge \Phi$ in the superpotential makes makes some components of $\Phi$ massive. Generally, since the Higgs field corresponds to the fluctuations of the 7-brane in the normal directions, this means that away from some special loci in the complex structure moduli of $B$, the 7-brane cannot ``wiggle" in specific ways. This can be seen more clearly, by noting that the coefficients of the spectral cover can be identified with the Casimirs of the Higgs field.

The above arguments were local, but in the following we will mainly focus one the effect of these deformations on the $G_4$ flux. In particular the type III example can only be explained in terms of the global $G_4$ flux stabilization rather than local two-flux.

\subsubsection*{Type I}\label{FtypeI}
We already know in the extension class of $V$ in \eref{extB} appeared as the new terms in $a_0$ and $a_2$ in \eref{extv2}. By the well-known arguments reviewed briefly above, these jumps in $a_0$ and $a_2$ corresponds to new umps in Casimirs of $\Phi$, in other words, new zero modes in fluctuations of the 7-brane. 

So new zero modes in the moduli space of the vector bundle in Heterotic string, corresponds to new zero modes in deformations of the 7-brane in F-theory.

To construct the $G_4$ flux, assume the spectral cover is smooth,
\begin{eqnarray}
    S= a_2 x+ a_0 z^2. \nonumber
\end{eqnarray}
Then by the introduction above, $G_4$ flux is related to the divisor $\gamma$ in $c_1(\mathcal{N})$. But $\gamma=0$ for $SU(2)$ bundles.\footnote{$\gamma$ induces non-zero chirality in the effective theory, but for $SU(2)$ bundles, the chirality must be zero. The vanishing of $\gamma$ can also be seen from direct spectral cover calculations.} So one may naively expect that the $G_4$ flux must vanish in this case. But, note that in this case, there are finite number of ``vertical components" (i.e. isolated elliptic curves) inside the spectral cover located above the points $a_0=a_2=0$ in the base $B$. In the dual F-theory geometry, such vertical components become 4-cycles $C_4$ by cylinder map which are (at least topologically) elliptic curves times a $\mathbb{P}^1$. 

One can use these cycles to construct $G_4$ flux,
\begin{eqnarray}
    G_4 \longrightarrow C_4 + \text{Corrections},
\end{eqnarray}
where the corrections are terms we need to add such that
\begin{eqnarray}
    G_4 \cdot \sigma \cdot \pi^*D_{base} =0.
\end{eqnarray}
Note that this $G_4$ flux only exists when the terms $a_2$ and $a_0$ in the spectral cover are not zero. Therefore, when we move to a generic point in the complex structure moduli where there are no jumps in $a_2$ and $a_0$, the $G_4$ flux constructed above ``disappears" i.e. $G_4$ becomes non-holomorphic, which by the Gukov-Vafa-Witten superpotential makes the corresponding complex structure deformation massive.\footnote{The Calabi-Yau four-fold of F-theory has $(4,6,12)$ singularity over the points of vertical components of $S$. Naively one may intend to blow these singularities up, but this is only possible when the Heterotic ``bundle" (coherent sheaf) is a direct sum of a vector bundle and small instantons  localized on the elliptic curves.}

\subsubsection*{Type II} \label{FtypeII}
In this case the spectral cover is smooth, and as explained in subsection \ref{typeII}, after the complex structure deformation the spectral cover is not algebraic anymore. So an argument similar to the one mentioned above is (when $a_2\ne 0$) also applies in this case. In other words, the dual Higgs bundle spectral data on the F-theory side is also required to be holomorphic, and these deformations are obstructed by the F-term equations in the 7-brane worldvolume theory.   

In addition, $\gamma$ term in \eref{specdata} can be non-zero in general (for $n>2$). The general form of $\gamma$ in \cite{Friedman:1997yq,Friedman:1997ih} is 
\begin{eqnarray}
\gamma = \lambda (n\sigma-\eta+n c_1(B))|_{S}.
\end{eqnarray}
The curves (divisors) $\sigma|_S$ and $\eta_S$, and hence $\gamma$ and $G_4$ in \eref{specdata}, are holomorphic as long as the spectral cover is holomorphic. Therefore, if the coefficients of the spectral cover depend on the new terms that appear in special loci in the complex structure moduli, then it is impossible to move away from those points since $\gamma$ and hence the corresponding $G_4$ flux becomes non-holomorphic.

In the particular example we considered in this paper $\gamma$ depends on the new divisor $e_1$,
\begin{eqnarray}
    \gamma=\gamma_0+\lambda_1 e_1.
\end{eqnarray}
Of course $\pi_{s*}\gamma \ne 0$ now, so one should modify the corresponding four-flux, but the point is that this four-flux is only holomorphic on the jumping locus.

\subsubsection*{Type III} \label{FtypeIII}
In this type the the $\gamma$ is modified as\footnote{Note that $\pi_{S*}\left(\frac{1}{n}\pi_*[C] -[C] \right)=0$.}
\begin{eqnarray}
\gamma = \lambda\left(n\sigma -\eta+n c_1(B) \right)+\left(\frac{1}{n}\pi_*[C] -[C] \right). 
\end{eqnarray}
As mentioned before on the Heterotic side the dependence of \eref{lineC} on $[C]$ stabilize both the complex structure moduli and the vector bundle moduli. But in F-theory dual both \eref{Fdef} and \eref{Sdef} corresponds to complex structure deformation on the F-theory side.

The corresponding (horizontal) four-flux that is responsible for stabilizing the F-theory complex structure is already studied \cite{Braun:2011zm}. A quick look at the defining relation for the F-theory geometry \eref{FthGeo} makes it clear that with this particular constraints over the geometry of the Heterotic Calabi-Yau and spectral cover, there is an algebraic four-cycle,
\begin{eqnarray}
C_4 : = \{X=Y=\alpha =0\} \subset X_F,
\end{eqnarray}
which can be used to construct an ``algebraic four-flux". It is also clear that $C_4$ is actually a $\mathbb{P}^1$ fibration (cylinder) over the algebraic curve $[C]$ in the Heterotic geometry. Therefore we can identify $C_4$ with the $P_{Cyl}^* [C]$ part of the four-flux induced by the spectral cover. 

Finally note that contrary to type I and II, the two-flux localised inside the 7-brane is not responsible for stabilizing the complex structure moduli. This is because the vector bundle in the Heterotic side is identified with the spectral data living in the \textit{compact elliptic fibration} over $B$, while the Higgs bundle on the 7-brane is identified with spectral data that live in the \textit{non-compact} fibration $Tot(K_B)$. Therefore in local F-theory/Heterotic duality only the part of Heterotic spectral data in the patch that contains the zero section is identified with the Higgs bundle spectral data, in other words,
\begin{eqnarray}
    (\mathcal{S}_{Het},\mathcal{N}_{Het})|_{\sigma\text{-patch}} \longleftrightarrow (\mathcal{S}_{Higgs},\mathcal{N}_{Higgs}).
\end{eqnarray}
However, remember that
\begin{eqnarray}
\sigma \cdot [C]=0.    
\end{eqnarray}

\section{Conclusion} \label{sec5}

In this paper, we have studied how the complex structure moduli of elliptically fibered Calabi-Yau threefold can be stabilized by holomorphic bundles in the context of a heterotic compactification.
Complex structure moduli of elliptically fibered Calabi-Yau include moduli from both the base and the fibration. 
We find three approaches 
In terms of spectral cover bundles, we find both of these moduli can be stabilized by the use of algebraic cycles. 
The idea is that on special locus in the moduli space of base (fibration), there exist emergent algebraic cycles. 
If one defines 
We develop efficient tools to identify the stabilized moduli and present concrete examples for all three approaches. 
As a result, we show that complex structure moduli can be effectively stabilized by bundle holomorphy.

As mentioned in the introduction, one of the advantages of working with spectral covers is the manifest connection with the F-theory and hence type II models. In other words, the complex structure moduli stabilization scenarios are already well known, at least conceptually, in both type II and Heterotic string, but apparently they look very different with each other. But both of these theories can be seen as different limits of F-theory. So by describing the moduli stabilization in terms of spectral data, we could see what should be the corresponding stabilization scenario in the dual F/type II theory. In particular we have shown that in the type I/II examples the dual stabilization is made possible by the flux inside the 7-brane worldvolume, while in the type III example the dual F-theory geometry is stabilized by a horizontal algebraic $G_4$ flux.

\section*{Acknowledgments}
W. Cui would like to thank James Gray for useful discussion. 

\appendix

\section{Complex Structure Moduli Space of Elliptically Fibered Calabi-Yau Threefold} \label{AppCS}

fibered
In this appendix we briefly discuss how the complex structure of the base $H^1(B,TB)$ is embedded into the complex structure of the Calabi-Yau threefold. We assume $B$ is a Fano surface ($K_B^{-1}$ is ample) and $h^1(\mathcal{O}_B)=0$.

One can use the adjunction sequence for the section, $\sigma$, of the fibration to relate the tangent bundle of the base $TB$ with the tangent bundle of the Calabi-Yau,
\begin{eqnarray}\label{KoszulAppC}
0\rightarrow TB \rightarrow TX|_B \rightarrow \mathcal{N}_{B/X}=K_B \rightarrow 0.
\end{eqnarray}
From the sequence above one can see, as long as the base $B$ is simply connected, $H^1(TB)\simeq H^1(B,TX|_B)$. So we should find the relation between $TX|_B$ and $TX$,
\begin{eqnarray}\label{adjunction}
0\rightarrow TX\otimes \mathcal{O}_X(-\sigma) \rightarrow TX \rightarrow TX|_B \rightarrow 0.
\end{eqnarray}
The corresponding long exact sequence is,
\begin{eqnarray}\label{longexactC3}
&0&\rightarrow H^0(TX\otimes \mathcal{O}_X(-\sigma)) \rightarrow H^0(TX) \rightarrow H^0(TX|_B) \rightarrow \nonumber \\
&&\rightarrow H^1(TX\otimes \mathcal{O}_X(-\sigma)) \rightarrow H^1(TX) \rightarrow H^1(TX|_B) \rightarrow \nonumber \\
&&\rightarrow H^2(TX\otimes \mathcal{O}_X(-\sigma)) \rightarrow H^2(TX) \rightarrow H^2(TX|_B) \rightarrow \nonumber \\
&&\rightarrow H^3(TX\otimes \mathcal{O}_X(-\sigma)) \rightarrow H^3(TX) \rightarrow 0.
\end{eqnarray}
We will show bellow that as long as $K_B^{-1}$ is ample, the homeomorphism $H^1(X,TX|_B)\simeq H^1(B,TB)\rightarrow H^2(TX\otimes \mathcal{O}_X(-\sigma))$ have to be zero. To see this we will use the following sequences,
\begin{eqnarray*}
&0\rightarrow TX\otimes \mathcal{O}_X(-\sigma) \rightarrow T\mathcal{A}|_X\otimes \mathcal{O}_X(-\sigma) \rightarrow \mathcal{N}|_X\otimes \mathcal{O}_X(-\sigma)\rightarrow 0,& \nonumber \\
&0\rightarrow T_{\bar{\pi}}|_X \rightarrow T\mathcal{A}|_X \rightarrow \pi^* TB \rightarrow 0,& \nonumber\\
&0\rightarrow \mathcal{O}_X \rightarrow \mathcal{O}_X(3\sigma-3K_B) \oplus \mathcal{O}_X(2\sigma-2K_B) \oplus \mathcal{O}_X(\sigma) \rightarrow T_{\bar{\pi}}|_X \rightarrow 0,&
\end{eqnarray*}
where $\bar{\pi}:\mathcal{A}\rightarrow B$ is the ambient space that the Calabi-Yau is embedded inside it.\footnote{Note that $\mathcal{A}$ is a $\mathbb{P}^{231}$ fibration.} Now, by the last sequence above, one can show 
\begin{eqnarray*}
\pi_* (T_{\bar{\pi}}|_X\otimes \mathcal{O}_X(-\sigma))=K_B^{-3}\otimes (K_B^{2}\oplus \mathcal{O}_B)\oplus K_B^{-2}\oplus \mathcal{O}_B.
\end{eqnarray*}
On the other hand, one can use the second sequence to show,
\begin{eqnarray*}
R\pi_*\left( T\mathcal{A}|_X\otimes\mathcal{O}_X(-\sigma)\right) = \pi_* (T_{\bar{\pi}}|_X\otimes \mathcal{O}_X(-\sigma)) \oplus TB\otimes K_B [-1].
\end{eqnarray*}
Finally the first short exact sequence, and the result above leads to,\footnote{Note that $\pi_* \left(TX\otimes \mathcal{O}_X(-\sigma) \right)=0$ because $TX$ is a stable bundle, and one can show its restriction on generic fiber of $X$ is given by non-trivial extension of trivial bundles. Therefore $TX\otimes \mathcal{O}_X(-\sigma)$ doesn't have any global section on generic fibers. Naively, we may say $\pi_* \left(TX\otimes \mathcal{O}_X(-\sigma) \right)$ should be a torsion sheaf on $B$ (probably non-zero only on the discriminant curve of $X$), but since $\pi$ is flat and $TX\otimes \mathcal{O}_X(-\sigma)$ locally free, the pushforward $\pi_*\left(TX\otimes \mathcal{O}_X(-\sigma) \right)$ must be locally free, and hence should vanish everywhere.}
\begin{eqnarray*}
0\rightarrow \pi_* (T_{\bar{\pi}}|_X\otimes \mathcal{O}_X(-\sigma)) \rightarrow (\pi_* \mathcal{N}|_X\otimes \mathcal{O}_X(-\sigma)) \rightarrow R^1\pi_* (TX\otimes \mathcal{O}_X(-\sigma))\rightarrow TB\otimes K_B\rightarrow 0,
\end{eqnarray*}
or, by considering the cokernel of the second map, we can rewrite this sequence as
\begin{eqnarray}\label{twistedTX}
0\rightarrow K_B^{-10}\otimes \mathcal{I}_{24K^2}\rightarrow R^1\pi_* (TX\otimes \mathcal{O}_X(-\sigma))\rightarrow TB\otimes K_B\rightarrow 0, 
\end{eqnarray}
where, $\mathcal{I}_{24K^2}$ is the ideal sheaf of a finite number of points given by the intersection of divisors $(-6 K_B)\cdot (-4K_B)$. Now we can talk about the cohomologies of $TX\otimes \mathcal{O}_X(-\sigma)$ in \eref{longexactC3}. By Leray spectral sequence,
\begin{eqnarray*}
E_2^{p,q}:=H^p(B,R^q\pi_* (TX\otimes \mathcal{O}_X(-\sigma))) \Rightarrow H^{p+q}(X,TX\otimes \mathcal{O}_X(-\sigma))).
\end{eqnarray*}
Since $R^0\pi_* (TX\otimes \mathcal{O}_X(-\sigma))=0$ then
\begin{eqnarray*}
H^p(X,TX\otimes \mathcal{O}_X(-\sigma)))=H^{p-1}(B,R^1\pi_* (TX\otimes \mathcal{O}_X(-\sigma))).
\end{eqnarray*}
The cohomology long exact sequence of the equation \eref{twistedTX} leads to,
\begin{eqnarray*}
&0&\rightarrow H^0(K_B^{-10}) \rightarrow H^0(R^1\pi_* (TX\otimes \mathcal{O}_X(-\sigma)))\rightarrow H^0(TB\otimes K_B)\rightarrow \nonumber \\
&&\rightarrow H^1(K_B^{-10}) \rightarrow H^1(R^1\pi_* (TX\otimes \mathcal{O}_X(-\sigma)))\rightarrow H^1(TB\otimes K_B)\rightarrow \nonumber \\
&&\rightarrow H^2(K_B^{-10}) \rightarrow H^2(R^1\pi_* (TX\otimes \mathcal{O}_X(-\sigma)))\rightarrow H^2(TB\otimes K_B)\rightarrow 0
\end{eqnarray*}
By the ampleness assumption of $K_B^{-1}$ we expect $H^{i}(B,K_B^{-10})=0$ for $i\ne 0$. So 
\begin{eqnarray*}
H^2(X,TX\otimes \mathcal{O}_X(-\sigma)))&=&H^1(R^1\pi_* (TX\otimes \mathcal{O}_X(-\sigma))) = H^1(TB\otimes K_B) = H^{1,1}(B),\\
H^3(X,TX\otimes \mathcal{O}_X(-\sigma)))&=& H^2(R^1\pi_* (TX\otimes \mathcal{O}_X(-\sigma))) = H^2(TB\otimes K_B) = H^{0}(B,TB^*).\nonumber
\end{eqnarray*}
At the end note that in general $h^2(X,TX)=h^{1,1}(X)=1+h^{1,1}(B)$, and also from \eref{KoszulAppC} we can show $H^2(TX|_B)=H^2(K_B)\oplus H^2(TB)=\mathbb{C}\oplus H^2(TB)$. Then to get the dimension of $H^2(TX)$ in \eref{longexactC3} correct, all of the elements of $H^2(X,TX\otimes \mathcal{O}_X(-\sigma)))=H^{1,1}(B)$ should inject into $H^2(TX)$. Therefore we get a surjection (at least as long as $H^1(\mathcal{O}_B)=0$),
\begin{eqnarray*}
H^1(TX)\rightarrow H^1(TX|_B)\simeq H^1(TB)\rightarrow 0.
\end{eqnarray*}
We conclude that the elements of $H^1(TB)$ corresponds to a subspace of the complex structure deformation of the Calabi-Yau $H^1(TX)$.

\section{Type II and Type II Fourier-Mukai Calculations} \label{AppFM}

For completeness, in this section, we present some of the Fourier-Mukai calculations' details as much as is necessary here. Following \cite{Anderson:2019agu} we can compute the Chern characters of the Fourier-Mukai $\Phi(V)$ of V. In type II example it would be,
\begin{eqnarray}
    Ch(\Phi(V))=-[\mathcal{S}]+\left([\mathcal{S}]\cdot \left(\frac{C_1(B)}{2}\right)+\frac{1}{2}c_3(V)[f]\right),
\end{eqnarray}
where $[\mathcal{S}]=\eta+n\sigma$. In type II example we chose $\eta=eta_0 + e_1$, and with this choice the algebraic equation of $[\mathcal{S}]$ will be reducible. Therefore the spectral line bundle $\mathcal{N}$ will be given as an extension, 
\begin{eqnarray}
    0\longrightarrow i_{e_1*} \mathcal{L}_1 \longrightarrow \mathcal{N}\longrightarrow i_{S_0*}\mathcal{L}_2 \longrightarrow 0,
\end{eqnarray}
where $i_{e_1}$ and $i_{S_0}$ are respectively the inclusion morphisms of the surfaces $\pi^* e_1$ and $S_0$ with divisor class $n\sigma+\eta_0$. Now one can use,
\begin{eqnarray}
\Phi(V)=i_{S*}\mathcal{N}[-1]\Longrightarrow Ch(\Phi(V))=-Ch(i_{S*}\mathcal{N})=-Ch(i_{e_1*} \mathcal{L}_1)-Ch(i_{S_0*}\mathcal{L}_2).
\end{eqnarray}
Next we can use the Grothendieck-Riemann-Rock to compute,
\begin{eqnarray}
Ch(i_{e_1*} \mathcal{L}_1) &=& [e_1]\cdot \left(\frac{e^{c_1(\mathcal{L}_1)}}{Td(\mathcal{O}(e_1))} \right), \\
Ch(i_{S_0*}\mathcal{L}_2) &=& [S_0]\cdot\left(\frac{s^{c_1(\mathcal{L}_2)}}{Td(\mathcal{O}(S_0))}\right).
\end{eqnarray}
An order by order comparison of the formulas written above leads to the following result,  
\begin{eqnarray}
c_1({\cal L}_2) &=& \frac{1}{2}(c_1(B)-[{\cal S}_0]) + \lambda \left(n\sigma - \eta_0 +n c_1(B) \right) + \lambda_1 e_1.
\end{eqnarray}
A similar calculation for the type III example would lead to,
\begin{eqnarray}
Ch(\Phi(V))=-[n\sigma+\eta+\pi^*\pi_*[C]]+[n\sigma+\eta+\pi^*\pi_*[C]]\cdot\left(\frac{c_1(B)}{2}-\frac{1}{2}\pi^*\pi_*[C] \right) + [C] +\dots.
\end{eqnarray}
Again order by order comparison of this with the Chern character of the spectral sheaf over a smooth surface $i_{S*}\mathcal{N}$ shows that first of all the divisor class of the spectral cover is
\begin{eqnarray}
[S]=n\sigma+\eta+\pi^*\pi_*[C],
\end{eqnarray}
second Chern class of the line bundle $\mathcal{N}$ must be,
\begin{eqnarray}
c_1(\mathcal{N}) &=& \frac{1}{2}\left([S(V_n)]-c_1(B)\right) + \lambda \left(n\sigma-\eta+n c_1(B)\right) +\left(\frac{1}{n}\pi_*[C] -[C] \right).
\end{eqnarray}

\section{Computation Details of ``Jumping'' Ext}\label{AppTypeI}

In this Appendix, we briefly outline the details of the calculation about the ``jumping'' of the line bundle cohomology and the number of moduli stabilized in Subsection \ref{typeI}. The base $B$ of the elliptic Calabi-Yau threefold $X$ is defined in (\ref{s3base}). 
The defining polynomial $p_0$ is an degree $(2,2)$ polynomial in $\{x_0,x_1\}$ and $\{y_0,y_1,y_2\}$. 
The coefficients of $p_0$ is a redundant description of the complex structure of $B$. We will extract the independent ones from them later.

The rank $2$ extension bundle $V$ is introduced in (\ref{extB}) with extension group $Ext(L,L^{\vee}) = H^1(X,L^2)$.
If the threefold $X$ admits an elliptic fibration, this group can be expressed as 
\begin{equation}
    H^1(X,L^2) = H^0(B, L_1\oplus L_2), 
\end{equation}
where $L_1 = O(D_b-C_1(B))$ and $L_2 = O(D_b+C_1(B))$ are line bundles and $D_b$ is a divisor on the base. 
Thus, the Ext group of $V$ is a direct sum of $H^0(B, L_1)$ and $H^0(B, L_2)$. 
We will study the ``jumping'' of both of them in the following.

The line bundle cohomology on the base $B$ follows from a similar procedure discussed in Subsection \ref{sec21}. 
As before, we can write the $L_1$ and $L_2$ in terms of the line bundles of the ambient space by Koszul sequence \cite{hartshorne}. 
The associated long exact sequence is 
\begin{equation} \label{koszul_long2}
0 \to H^0(B, L) \to
H^1({\cal A}, N^{\vee}\otimes L_{{\cal A}})
\stackrel{p_{0}}{\rightarrow} H^1({\cal A}, L_{{\cal A}}) \to
H^1(B, L) \to 0 \ .
\end{equation}
Notice that, instead of $H^1(X, L^2)$, we are interested in the $H^0(B, L)$ here. 
For our example considered in Section \ref{sec3}, $L_1=O(-2,3)$ and $L_2=O(-2,5)$. 
From the equation (\ref{koszul_long2}), we find 
\begin{align} \label{Appkernel}
    H^0(X,L_1) &= \textnormal{Ker} \left[ H^1(A,O(-4,1)) \stackrel{p_0}{\rightarrow} H^1(A,O(-2,3)) \right] \\
    H^0(X,L_2) &= \textnormal{Ker} \left[ H^1(A,O(-4,3)) \stackrel{p_0}{\rightarrow} H^1(A,O(-2,5)) \right]
\end{align}
To determine the kernel, we express the cohomology of the source and the target in terms of their Bott-Borel-Weil polynomial representations. For the first one in Equation (\ref{Appkernel}), they are 
\begin{align} \label{bbw}
H^1(\mathbb{P}^1 \times \mathbb{P}^2,O(-4,1)) & = \{ \frac{y_{0}}{x_{0}^{2}},  \frac{y_{1}}{x_{0}^{2}},\ldots  \frac{y_{0}}{x_{0}x_{1}}, \ldots \frac{y_{2}}{x_{1}^{2}} \} \\
H^1(\mathbb{P}^1 \times \mathbb{P}^2,O(-2,3)) & =\{ y_{0}^{3}, y_{0}^{2} y_1, \ldots y_{2}^{3}\} 
\end{align}
and $p_0$ is the degree $(2,2)$ defining polynomial. Note that the multiplication of $\frac{y_{0}}{x_{0}x_{1}}$ and $x_0^2y_0^2$ maps to zero in the target because the resulting monomials have the wrong degree and should not be considered as an appropriate element of the target. The same is for a map like $x_1^2y_0^2$. Thus, the most general form of the $p_0$ that have non-trivial kernel is given by 
\begin{equation} \label{specialCS}
    p_0 = x_1^2P_1(y)+x_2^2P_2(y),
\end{equation}
where $P_1(y)$ and $P_2(y)$ are arbitrary degree $2$ polynomials in $\{y_0,y_1,y_2\}$. 
There are three independent monomials with the denominator $x_{0}x_{1}$. Therefore, the dimension of the kernel ``jumps'' to $3$ when the defining polynomial has the form in (\ref{specialCS}) and it vanishes when $p_0$ is a generic polynomial. By the same approach, we also compute the ``jumping'' of $H^0(X,L_2)$. The results are summarized in Equation (\ref{h0jump}) that is used for complex structure moduli stabilization in Section \ref{sec3}.

\section{Approach for Identifying Reducible Deformations}\label{appT2}

In Subsection \ref{typeII}, we find that the space of reducible deformations $M^2_{\;\;r}$ is generated by $5$ out of $6$ polynomials given in (\ref{es6}).
There, we use the fact that polynomial deformation $\delta F_2$, if reducible, is equivalent to shifts of the coefficients in $F^{(0)}_2$. 
We will see why it is true and verify that $M^2_{\;\;r}$ contains all the possible reducible deformations. In addition, we will generalize the method introduced in Subsection \ref{typeII} to find reducible deformations of other initial polynomials.

Let's expand the initial $F^{(0)}_{2}$ defined in (\ref{reducibleF}) as 
\begin{eqnarray} \label{irredunF}
F^{(0)}_{2} = (a u_1^2 + b u_1u_2 + c u^2_2) z_1^2 + (d u_1 + e u_2) z_1z_2 + fz^2_2 \;.
\end{eqnarray}
where coefficients $a,b,c,d,e,f$ are given by 
\begin{eqnarray} \label{condR}
    &a  = A_1 B_1 \;, \qquad &b  = A_1 B_2 + A_2 B_1 \;, \quad c  = A_2 B_2 \;, \\ \nonumber
    &d  = A_1 B_3 + A_3 B_1 \;, \quad &e  = A_2 B_3 + A_3 B_2 \;, \quad f  = A_3 B_3\;
\end{eqnarray} 
with $A_1,A_2,A_3,B_1,B_2,B_3$ the coefficient of $F^{(0)}_2$ in the factored form. 
It means that given a polynomial with coefficients $a,b,c,d,e,f$, if one can find a set of $\{A_i\}$ and $\{B_i\}$ solves the equation (\ref{condR}), the polynomial is reducible. This is the reducible condition that will be used later. 
Obviously, $F^{(0)}_{2}$ is expressed by the following $6$ monomials
\begin{equation} \label{mono_basis}
     (u_1^2z_1^2, u_1u_2z_1^2, u^2_2 z_1^2, u_1z_1z_2, u_2z_1z_2, z^2_2)\;.
\end{equation}
Thus, there are at most $6$ polynomial deformations for $F^{(0)}_{2}$. Each of those is given by the deformation of a monomial. We will study which of these monomial deformation or their combinations $\delta F_2$ can keep $F^{(0)}_{2}+\delta F_2$ reducible.

Suppose we deform the initial $F^{(0)}_2$ by the first monomial $\delta a u_1^2z_1^2$. 
The coefficients of the deformed polynomial $F^{(0)}_2 + \delta a u_1^2z_1^2$ becomes $\{a+\delta a,b,c,d,e,f\}$, which does not satisfy the reducible condition in (\ref{condR}). Thus, the deformed polynomial is not reducible.
To restore the reducibility, one needs to adjust the coefficients $\{A_i\}$ and $\{B_i\}$ so that they give rise to the new parameters $\{a+\delta a,b,c,d,e,f\}$ via the condition (\ref{condR}). 
Since $a = A_1 B_1$, it can be done by simply redefining $A_1$ to be $A'_1 = A_1 + \frac{\delta a}{B_1}$. Now, $a + \delta a = A'_1 B_1$. To satisfy the reducible condition (\ref{condR}), one needs to replace the $A_1$ with $A'_1$ in all other equations involving $A_1$, which are $b = A'_1 B_2 + A_2 B_1$ and $d = A'_1 B_3 + A_3 B_1$. 
The redefinition is equivalent to shift $b \to b+(\frac{B_2}{B_1})\delta a $ and $d \to d+(\frac{B_3}{B_1})\delta a$. That means, to preserve the reducibility, whenever we deform the $F$ by $\delta a u_1^2z_1^2$, we have to add the extra terms $(\frac{B_2}{B_1})u_1u_2z_1^2\delta a $ and $(\frac{B_3}{B_1})u_1z_1z_2 \delta a $ as well. The reducible deformation after a redefinition of $A_1$ is
\begin{equation}
    e_1 = u_1^2z_1^2 + (\frac{B_2}{B_1})u_1u_2z_1^2 + (\frac{B_3}{B_1})u_1z_1z_2, 
\end{equation}
which, by a factor of $B_1$, is the same deformations $e^A_1 = B_1 e_1$ as the one obtained by a shift of $A_1$ in (\ref{es6}).

Besides the redefinition of $A_1$, to absorb the deformation $\delta a u_1^2z_1^2$, one can also consider the redefinition of the $B_1$ as $B'_1 = B_1 + \frac{\delta a}{A_1}$. From the relation (\ref{condR}), one has to redefine the $b$ and $d$ by shifting $b \to b+(\frac{A_2}{A_1})\delta a $ and $d \to d+(\frac{A_3}{A_1})\delta a$, which is given by 
\begin{equation}
    e_2= u_1^2z_1^2 + (\frac{A_2}{A_1})u_1u_2z_1^2 + (\frac{A_3}{A_1})u_1z_1z_2.
\end{equation}
Again, up to a factor of $A_1$, it is the same deformation obtained by a shift of $B_1$. Following the same analysis, we consider all $6$ monomial deformations in (\ref{mono_basis}). We find that all these deformations can be absorbed by the shifts of the coefficients $\{A_i\}$ and $\{B_i\}$, which gives the same reducible deformations as ones in $M^2_{\;\;r}$.

The reducible deformations considered above are obtained by the redefinition of one coefficients. For example, $e_1$ and $e_2$ are generated by a shift of $A_1$ and $B_1$. One may suspect that can we get more reducible deformations by 
redefining $A_1$ and $B_1$ at the same time? For example, given the monomial deformation $\delta a u_1^2z_1^2$, one can absorb it by shifting the $A_1$ and $B_1$ at the same time by $\delta A_1 = \frac{\delta a}{2B_1}$ and $\delta B_1 = \frac{\delta a}{2A_1}$. It is easy to check that, to first-order, the reducible condition (\ref{condR}) is satisfied. In particular, $a+\delta a = (A_1 + \delta A_1)(B_1 + \delta B_1)$. The corresponding reducible deformations $E$ follow by shifting the $A_1$ and $B_1$ in all other equations in (\ref{condR}).   
One can show that $E = \frac{1}{2}e_1+\frac{1}{2}e_2$, which means that $E$ is included in $M^2_{\;\;r}$ and redefining multiple coefficients at the same time does not give new reducible deformations. 
Therefore, we verify that $M^2_{\;\;r}$ contains all the possible reducible deformations and they are generated by the shifts of each coefficient in (\ref{reducibleF}) independently.

We generalize the above method to study the reducible deformations of polynomial other than (\ref{tunedP}). 
Suppose the initial $F_2^{(0)}$ is a product of a polynomial with $m$ terms and a polynomial with $n$ terms 
\begin{equation}
    F_2^{(0)} = (A_1 m^A_1 + A_2 m^A_2+\ldots +A_{r} m^A_{r})(B_1 m^B_1+B_2 m^B_2+\ldots +B_s m^B_s)
\end{equation}
where $\{m^A_i\}, i=1,2,\ldots, r$ and $\{m^B_j\}, j=1,2,\ldots, s$ are monomials and $\{A_i\}$ and $\{B_j\}$ are the corresponding coefficients. As before, the reducible deformations can be obtained by shifts of each coefficient in $\{A_i\}$ and $\{B_j\}$, 
\begin{eqnarray} \label{genBasis}
    &e^A_1 =  (B_1 m^B_1+B_2 m^B_2+\ldots +B_s m^B_s)m^A_1, \quad 
    &e^B_1 = (A_1 m^A_1 + A_2 m^A_2+\ldots +A_{r} m^A_{r})m^B_1  \\ \nonumber
    &e^A_2 =  (B_1 m^B_1+B_2 m^B_2+\ldots +B_s m^B_s)m^A_2, \quad 
    &e^B_2 = (A_1 m^A_1 + A_2 m^A_2+\ldots +A_{r} m^A_{r})m^B_2,  \\ \nonumber
    &\qquad \vdots  &\hspace{4cm} \vdots\\
    &e^A_r =  (B_1 m^B_1+B_2 m^B_2+\ldots +B_s m^B_s)m^A_r, \quad 
    &e^B_s = (A_1 m^A_1 + A_2 m^A_2+\ldots +A_{r} m^A_{r})m^B_s \;.  \nonumber
\end{eqnarray}
There are totally $(r+s)$ such polynomials. As the equation (\ref{relationR}), one can verify that these deformations satisfy the relation
\begin{equation} \label{genRelation}
F_2^{(0)} + \left(A_1 e^A_1 + A_2 e^A_2 + \ldots + A_r e^A_r \right) = F_2^{(0)} + \left(B_1 e^B_1 + B_2 e^B_2 + \ldots + B_s e^B_s \right) \;.
\end{equation}
Thus, there are $(r+s-1)$ independent polynomials and they form a basis of $M^2_{\;\;r}$.

\end{document}